\documentclass[3p,final,12pt]{elsarticle}

\usepackage{graphicx}
\usepackage{epstopdf}
\usepackage{float}
\usepackage{booktabs}
\usepackage{caption}
\usepackage{subcaption}
\usepackage{natbib}
\usepackage{tabularx}
\usepackage{lipsum}  
\usepackage{amsmath}
\usepackage{amsthm}

\usepackage{amssymb}
\usepackage[symbol]{footmisc}
\usepackage{hyperref}
\usepackage{color,soul}

\begin{document}

\begin{frontmatter}



\title{Computational modeling of Pulsed Field Ablation for pulmonary vein isolation}


\author[inst1]{Ashkan Bagherzadeh}

\author[inst3,inst4]{Nagib T Chalfoun}

\author[inst1,inst2,inst5]{Tong Gao}

\author[inst1]{Lik Chuan Lee}

\affiliation[inst1]{organization={Department of Mechanical Engineering, Michigan State University},
            addressline={428 S. Shaw Lane}, 
            city={East Lansing},
            postcode={48824}, 
            state={MI},
            country={USA}}

\affiliation[inst2]{organization={ Department of Computational Mathematics, Science and Engineering, Michigan State University},
            addressline={428 S. Shaw Lane}, 
            city={East Lansing},
            postcode={48824}, 
            state={MI},
            country={USA}}

\affiliation[inst3]{organization={Department of Medicine, Michigan State University},
            %addressline={428 S. Shaw Lane}, 
            city={East Lansing},
            postcode={48824}, 
            state={MI},
            country={USA}}

\affiliation[inst4]{organization={Corewell Health},
            %addressline={428 S. Shaw Lane}, 
            city={Grand Rapids},
            postcode={49503}, 
            state={MI},
            country={USA}}

\affiliation[inst5]{organization={Department of Mechanical Engineering, Tufts University},
            addressline={200 College Avenue}, 
            city={Medford},
            postcode={02155}, 
            state={MA},
            country={USA}}

%\author[inst1,inst2]{Author Three}

%\affiliation[inst2]{organization={Department Two},%Department and Organization
%            addressline={Address Two}, 
%            city={City Two},
%            postcode={22222}, 
%            state={State Two},
%            country={Country Two}}

\begin{abstract}
Pulsed field ablation (PFA) has emerged as a non-thermal alternative to traditional thermal ablation techniques for the treatment of atrial fibrillation (AF). This study presents a patient-specific 3D computational framework to model the effects of PFA on pulmonary vein isolation (PVI). The modeling framework is rigorously validated against published numerical and experimental data, demonstrating strong agreement across a range of scenarios. Using realistic left atrial (LA) anatomy, commercially available circular, flower, and basket catheter configurations are simulated to evaluate lesion formation across different applied voltages. The performance of each catheter type is quantitatively assessed using multiple metrics, including lesion volume, energy delivery efficiency and transmurality. Simulation results show that circular catheters provide the highest energy delivery efficiency and target coverage at lower voltages, while basket catheters produce the largest lesion volumes. This framework offers a useful basis for exploring catheter design and treatment planning in PFA applications.

\end{abstract}




\end{frontmatter}

\section{Introduction}
\noindent Atrial fibrillation (AF) is the most common form of arrhythmia that is caused by disorganized electrical impulses in the atria, which leads to irregular and rapid heart rates \cite{sagris2021atrial}. This condition significantly increases the risk of stroke and can contribute to the development of heart failure \cite{schnabel2024early}. Catheter ablation is known as an effective and minimally invasive treatment method for controlling AF, especially in patients with persistent symptoms or those who do not respond well to medications \cite{berenfeld2022atrial}. Targeting primarily the pulmonary veins where abnormal electrical signals often initiate AF, the procedure has been shown to improve symptoms and reduce arrhythmia recurrence \cite{becher2024atrial}. 
Catheter ablation for AF can be performed using different techniques, each with its own approach of targeting and isolating the sources triggering the arrhythmia. The most widely used techniques for AF ablation are radiofrequency ablation (RFA) \cite{hong2020catheter} and cryoballoon ablation \cite{tzeis20242024}, both utilizing thermal energy to ablate tissue. In recent years however, pulsed field ablation (PFA) has emerged as a promising alternative, where high voltage electric pulses are applied to selectively ablate tissue with minimal thermal injury and collateral damage \cite{virk2019catheter}.

In PFA, a catheter is introduced into the heart through a vascular access point, typically via the femoral vein. The catheter is guided to the left atrium with the help of advanced imaging techniques such as fluoroscopy and intracardiac echocardiography. Once it is positioned near the pulmonary veins or other targeted locations, the electrodes are deployed in various geometric configurations to deliver electric pulses. These pulses generate strong electric fields that disrupt cell membranes in the targeted area through a process called electroporation that causes irreversible damage to arrhythmogenic cells while sparing the adjacent tissues, including the nerves, esophagus, and coronary arteries. The non-thermal nature of PFA minimizes the risk of complications such as pulmonary stenosis or collateral injury, making it a promising approach for improving the safety and efficacy of AF ablation.

\begin{figure}[H]
    \centering
        \includegraphics[width=0.75\textwidth]{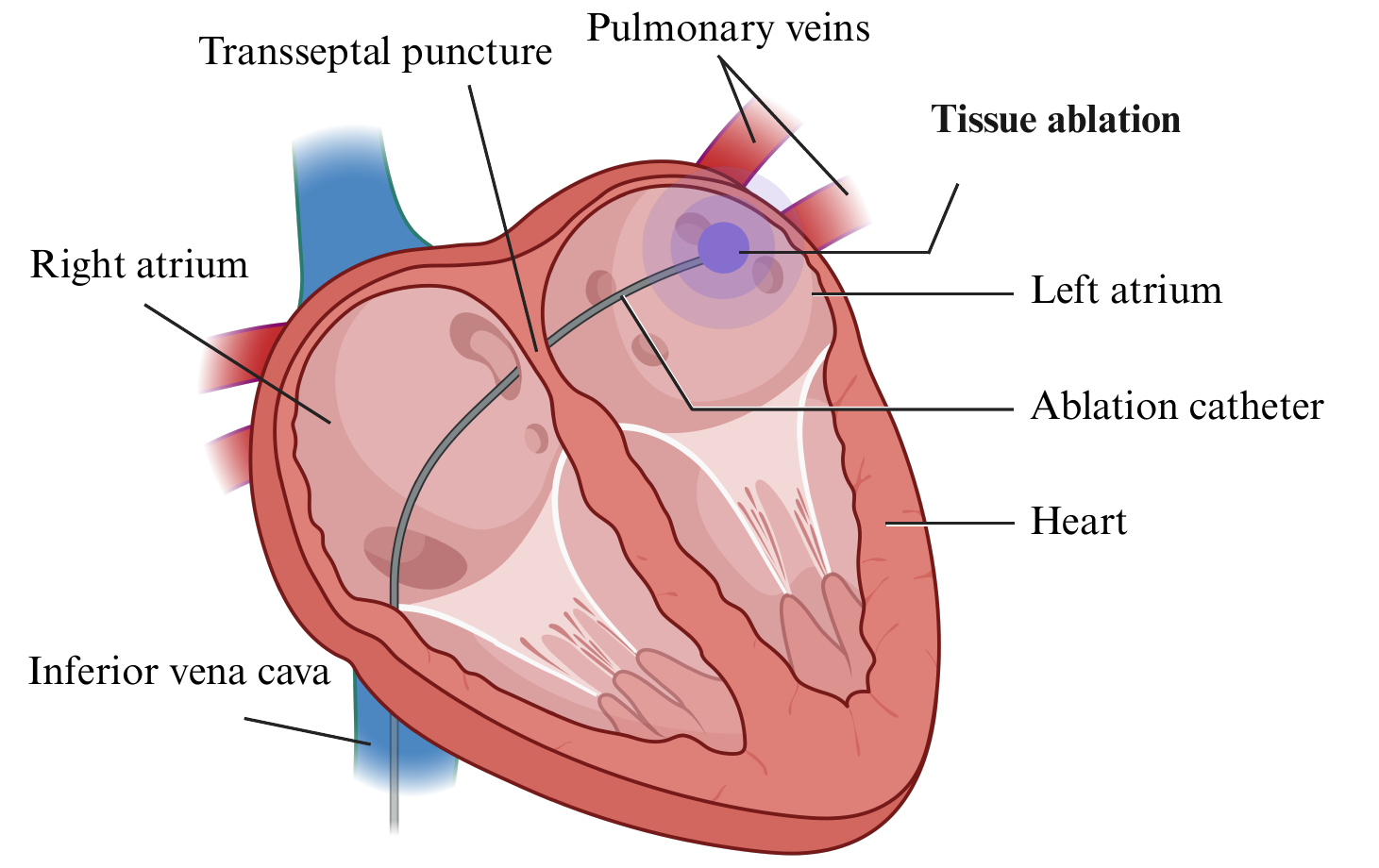}
        \label{figa_catheter}

    \caption{Schematic of PFA targeting pulmonary vein tissues and the catheter’s path into the left atrium}
    \label{pfa}
\end{figure}

Numerical simulations can play an important role in advancing our understanding and optimization of PFA by offering insights into electric field distribution, lesion predictability, and treatment planning that are difficult to obtain through experimentation alone. Most computational PFA models focus on the tissue or organ scale, assuming homogeneous properties and employing macroscopic electroporation thresholds. Some efforts, such as Krassowska and Filev’s single-cell model \cite{krassowska2007modeling}, explore the process at the cellular level, simulating transmembrane voltage dynamics and pore formation. However, clinical decision-making and device optimization typically depend on tissue-scale predictions.
At the tissue level, simulation studies have demonstrated how lesion characteristics depend on variables like voltage amplitude and catheter placement. \citet{Meckes2022} showed nonlinear relationships between applied voltage and lesion depth, while \citet{Gomez2022} reported that PFA lesions are more symmetric and less affected by blood flow compared to thermal methods. \citet{ji2022effect} found that sequential pulsing in multielectrode catheters enhances transmural lesion continuity. These findings underscore the utility of computational models in guiding energy delivery protocols and catheter design.
Several prior studies have examined the influence of catheter geometry and configuration on electric field patterns. \citet{Marino2021} an open-source simulation platform (openEP) for modeling electroporation with variable tissue and pulse parameters, although its main applications focus on electrochemotherapy. Belalcazar and colleagues \cite{belalcazar2021safety,belalcazar2024comparison} explored how minor differences in electrode positioning or catheter shape can significantly affect lesion predictability, highlighting the sensitivity of outcomes to design parameters.
While simulations provide powerful tools for predicting electric field behavior and guiding PFA design, experimental validation remains essential to confirm their biological relevance. \citet{Yao2017} demonstrated that combining high- and low-voltage pulses significantly enlarges ablation zones and induces measurable thermal effects, highlighting the importance of considering pulse protocol design. Preclinical work by \citet{Kawamura2022} in porcine models confirmed that lesion characteristics vary with catheter orientation and pulse repetition—findings consistent with prior numerical studies. \citet{howard2022effects} highlighted the critical role of electrode-tissue proximity, showing that even small offsets can markedly alter lesion formation. \citet{Arena2012} further explored these dynamics in a controlled in vitro setup using a 3D tumor model, quantifying electroporation thresholds in tissue-mimicking materials.

The aim of this work is to develop a 3D patient-specific computational modeling framework with anatomically realistic geometries to simulate PFA in clinically relevant scenarios. Particular attention is given to modeling commercially available catheter designs—circular, flower, and basket configurations—with detailed electrode geometries derived from manufacturer specifications to closely match clinical practice. This study also systematically investigates how catheter design and configuration affect performance under similar conditions using different metrics. The framework is rigorously validated against previously published experimental and computational benchmarks, ensuring reliable predictions across various scenarios. This computational framework establishes a foundation for patient-specific PFA simulations in future research. The structure of the paper is as follows: Section 2 outlines the computational framework, encompassing the mathematical formulation of the problem and the geometry. Section 3 focuses on the validation process and the simulation results. In Section 4, the findings are discussed in detail. Finally, Section 5 provides the concluding remarks of the study.

\label{intro}

\section{Methods}
\noindent Here, we provide an overview of the left atrium (LA) and pulmonary vein geometries as well as catheter design and specifications utilized in this study.  Governing equations, thermophysical properties, and simulation settings are also outlined here.

\subsection{Left atrium geometry}

\begin{figure}[H]
    \centering
        \centering
        \includegraphics[width=0.9\textwidth]{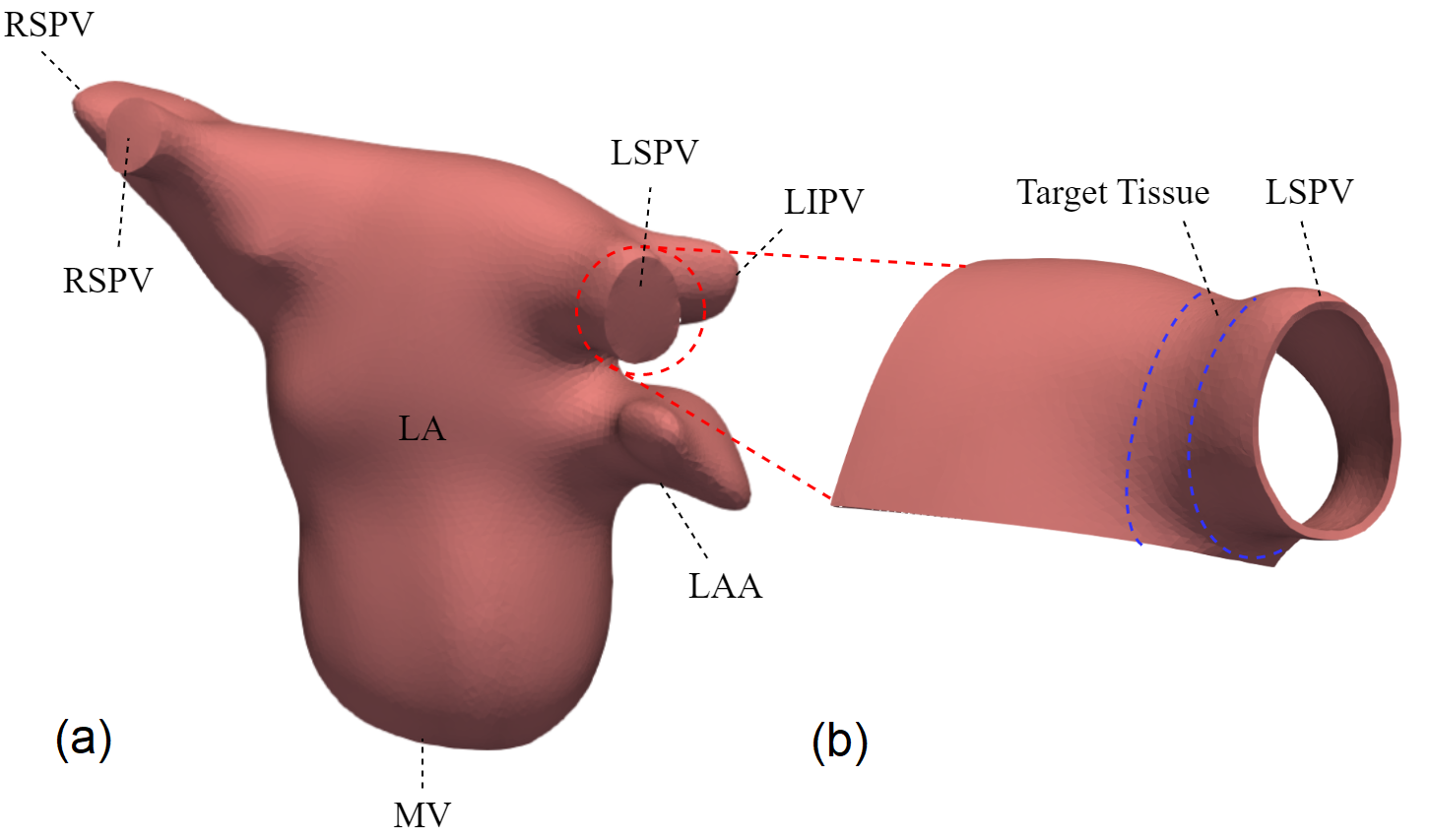}
    \caption{(a) LA geometry with key labeled anatomical features. (b) Simulation region consisting of LSPV and the ablation target at the vein ostia} 
        \label{LA-geo}
\end{figure}

\noindent Figure \ref{LA-geo} shows the geometry of LA with key anatomical features, which are namely, the left superior pulmonary vein (LSPV), left inferior pulmonary vein (LIPV), right superior pulmonary vein (RSPV), right inferior pulmonary vein (RIPV), left atrial appendage (LAA), and the mitral valve (MV) opening. The simulation domain consists of a region of the LA extending to the LSPV, which has an inlet diameter $\sim$14.5 mm and a wall thickness of 1 mm. The ablation target is located in LSPV ostia (see region defined by the blue-dotted line) and has a total volume of 385~mm\textsuperscript{3}.

\subsection{Catheter Geometry and Configurations}

\begin{figure}[H]
    \centering
    \begin{subfigure}[b]{0.33\textwidth}
        \centering
        \includegraphics[width=\textwidth]{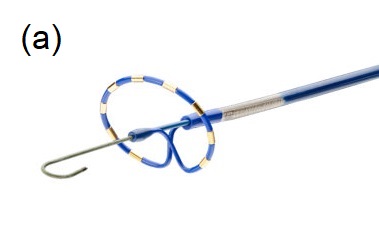}
    \end{subfigure}%
    \begin{subfigure}[b]{0.33\textwidth}
        \centering
        \includegraphics[width=\textwidth]{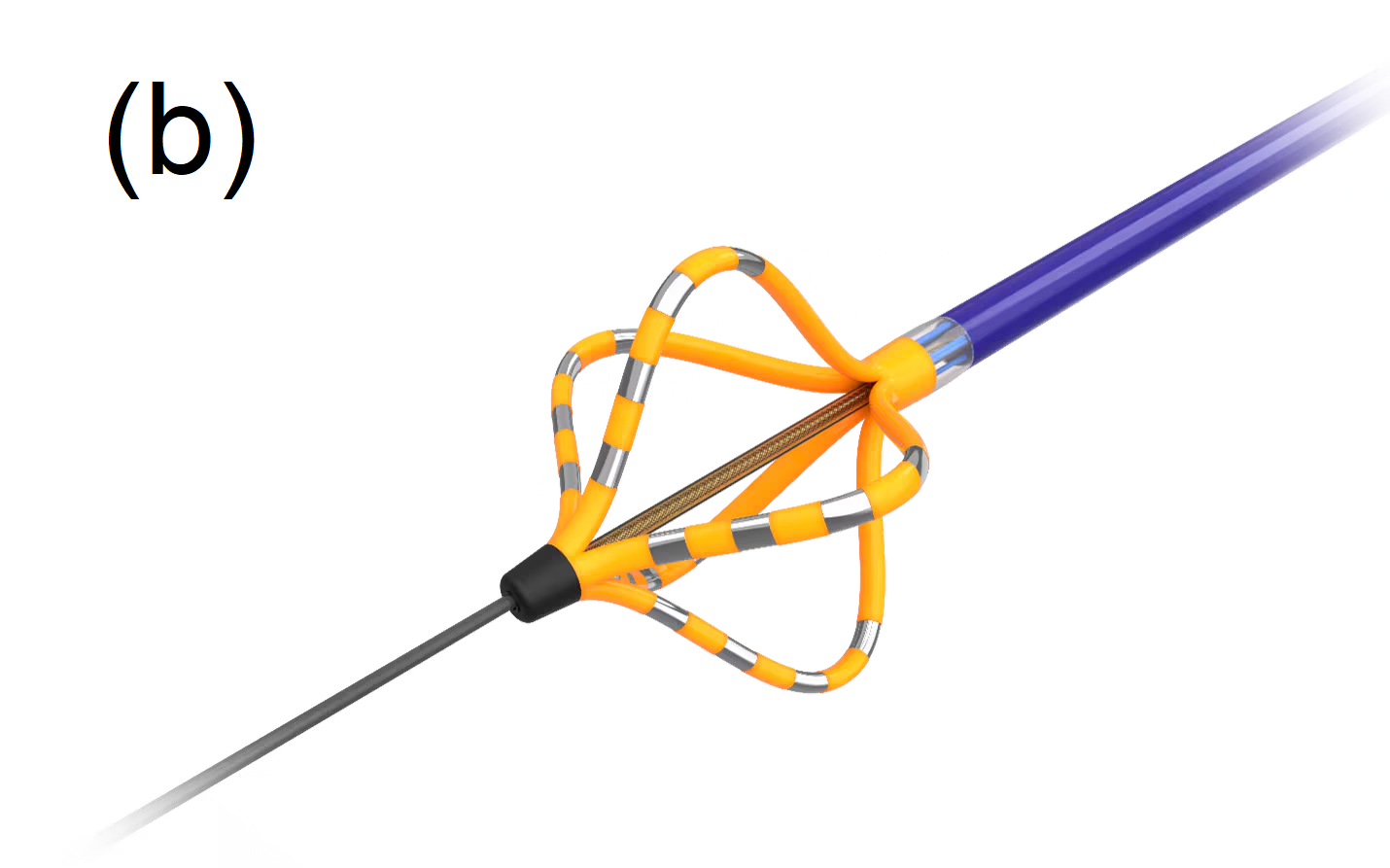}
    \end{subfigure}%
    \begin{subfigure}[b]{0.33\textwidth}
        \centering
        \includegraphics[width=\textwidth]{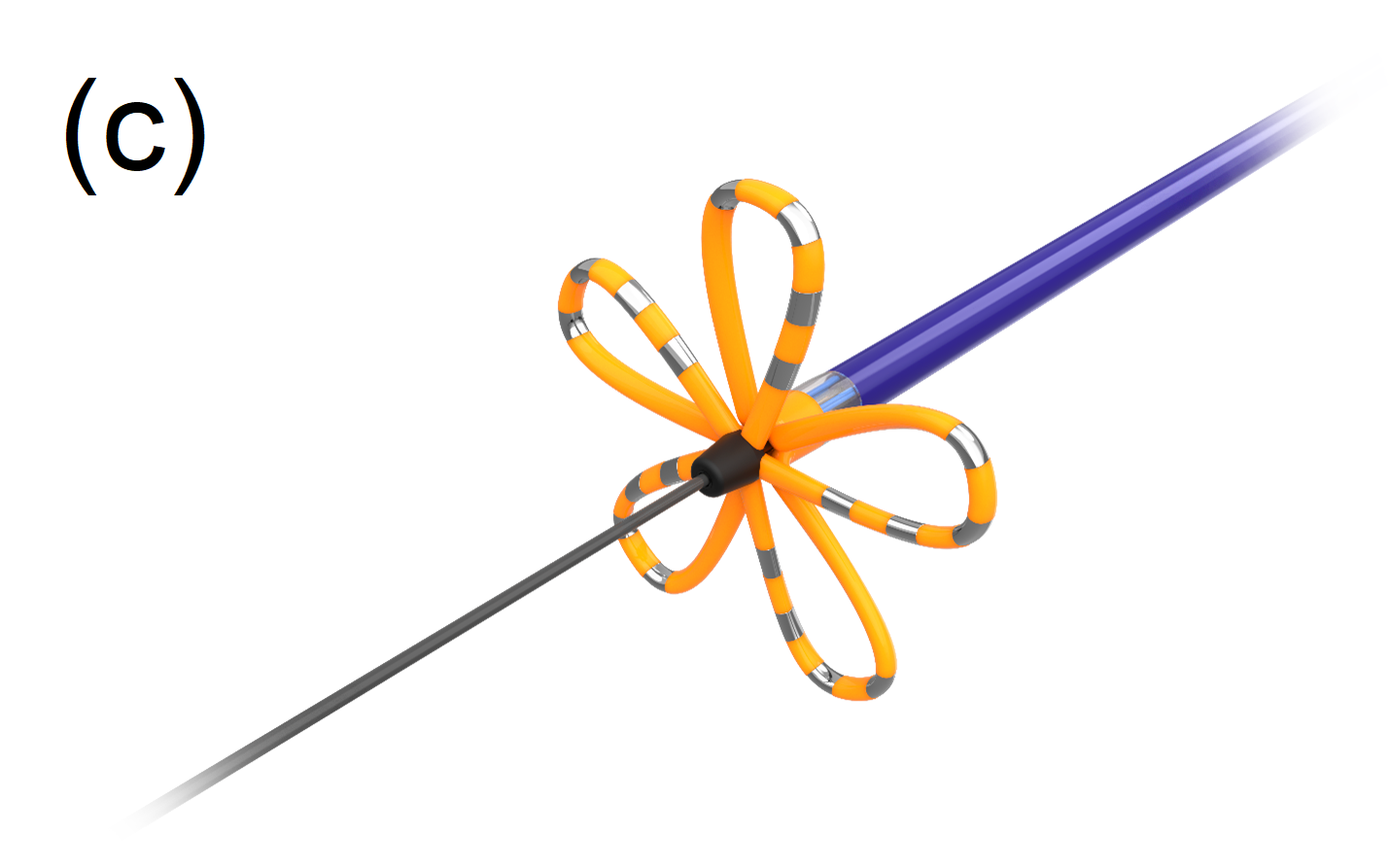}
    \end{subfigure}%
    \caption{(a) Circular,  (b) basket  and  (c) flower catheter configurations}
    \label{cath-geo}
\end{figure}

\noindent We consider 3 different types of catheter configurations from Boston Scientific (FARAWAVE\texttrademark) and Medtronic (PulseSelect\texttrademark) based on specifications provided in their respective data sheets \cite{farawave_datasheet,pulseselect_datasheet}. They are namely, circular, basket and flower catheter configurations (Figure \ref{cath-geo}).
In the basket configuration, the catheter maintains a partially open shape to facilitate safe positioning and preliminary contact with the tissue; once placed, it can expand into the flower configuration, where the splines splay outward like petals to enhance contact and ablation. The dimensional specifications of these catheters are given in Table \ref{tab:catheter_characteristics}.

\begin{table}[H]
\centering
\begin{tabular}{lccc}
\hline
\textbf{Metric}                      & \textbf{Basket} & \textbf{Flower} & \textbf{Circular} \\ \hline
Number of Electrodes                      & 20              & 20              & 9       \\ 

Each Electrode Surface Area (mm²)         & 50.24             & 50.24            & 15.46       \\ 
Electrode Length (mm)        & 3.25          & 3.25          & 3             \\
Inter-Electrode Distance (mm)        & 4.5          & 4.5          & 3.75             \\
Max. Diameter (mm)                   & 27         & 31          & 25                \\ \hline
\end{tabular}
\caption{Characteristics and dimensions of the Basket, Flower, and Circular Catheters}
\label{tab:catheter_characteristics}
\end{table}

\subsection{Governing Equations}
\label{methods}
\noindent Electric potential field \( \phi \) in both the blood and LA wall is governed by the Laplace equation under electrostatic conditions

\begin{equation}
\nabla \cdot (\sigma \nabla \phi) = 0 \hspace{2pt},
\end{equation}

\noindent where \( \phi \) is the electric potential, and \( \sigma \) is the electrical conductivity, which differs between blood and tissue. The electrical conductivity also varies with temperature \( T \) by 
\begin{equation}
\sigma(T) = \sigma_0 \cdot f(T) =   \sigma_0 (1+\alpha (T - T_b)) \hspace{2pt},
\label{eq:conductivity_relationship}
\end{equation}

\noindent In Eq. \ref{eq:conductivity_relationship},  \( \sigma_0 \) is the baseline conductivity, and \( f(T) \) is a linear function with constant \( \alpha \).

%\LC what are the coefficients? explained
\noindent Temperature distribution \( T \) in the tissue is governed by the Pennes Bioheat Equation 

\begin{equation}
\rho_t c_p \frac{\partial T}{\partial t} = \nabla \cdot (k_t \nabla T) + Q_{\text{elec}} - \rho_b c_b \omega_b (T - T_b) \hspace{2pt},
\end{equation}

\noindent where \( \rho_t \) , \( c_p \) and \( k_t \)  denote the tissue density, tissue specific heat capacity and tissue's thermal conductivity, respectively. 
Additionally, \( \rho_b \) , \( c_b \), \( \omega_b \)  and \( T_b \) denote the density, specific heat capacity, perfusion rate and temperature of the blood, respectively. 
The term \( Q_{\text{elec}} \), which accounts for the heat generated by the electric field \( \mathbf{E} \) $= - \nabla \phi$  through Joule heating, is defined as

\begin{equation}
Q_{\text{elec}} = \sigma |\mathbf{E}|^2 \hspace{2pt}.
\end{equation}

\noindent Ablated region where irreversible electroporation occurs in the LA wall is determined by an electric field threshold $|\mathbf{E}| \geq  268 \text{V/cm (or 26.8 V/mm)}$.
This value is based on {\it in vivo} lesion depth data measured in animal in a previous study \citet{Meckes2022}. 
These equations are solved using the finite element method (FEM) implemented using FEniCS \cite{alnaes2015fenics,logg2012automated}, and the key simulation parameters are summarized in Table \ref{tab:par}.
% LC: For completeness, you may want to put the weak form in the Appendix and refer to them.

\begin{table}[h!]
\centering
\begin{tabular}{lll}
\hline
\textbf{Parameter}      & \textbf{Value}     & \textbf{Unit} \\ \hline

Tissue density \( \rho_t \)         & 1000               & kg/m³ \\ 
Tissue specific heat capacity \( c_p \) & 4000          & J/(kg·K) \\ 
Tissue thermal conductivity \( k_t \) & 0.55              & W/(m·K) \\ 
Blood density \( \rho_b \)        & 1060               & kg/m³ \\ 
Blood specific heat capacity \( c_b \) & 3900          & J/(kg·K) \\ 
Blood perfusion rate \( \omega_b \) & 0.01            & 1/s \\ 
Blood temperature \( T_b \)       & 310                & K \\ 
Electric field threshold \( E_{\text{threshold}} \) & 26.8 & V/mm \\ 
Baseline Tissue electrical conductivity \( (\sigma_0)_t \) & 0.7              & S/m \\
Baseline Blood electrical conductivity \( (\sigma_0)_b \) & 0.4              & S/m \\
Baseline Tissue electrical conductivity \( \alpha \) & 0.005              & 1/K \\\hline
\end{tabular}
\caption{Parameters and their typical values for simulations}
\label{tab:par}
\end{table}

\subsection{Simulation Cases}

\noindent  We consider the following simulation cases to validate the computational model against numerical and experimental predictions  (Cases 1 -- 4). Thereafter, we simulate and compare the performance of different catheter configurations in ablating the PV ostium of the LA geometry in Figure \ref{LA-geo}  (Case 5).

\subsubsection*{Case 1: Electric field distribution validation}
\noindent The solver’s accuracy in predicting the electric field distribution was evaluated in this case. The computational model was set up to replicate the conditions of Meckes’ simulations \cite{Meckes2022}. A two-electrode catheter configuration was modeled in a blood pool adjacent to the endocardium, simulating an {\it in vivo} intracardiac environment. The study examined applied voltages ranging from 100 V to 2500 V. The catheter was tested at different distances from the endocardial surface (ranging from 0 mm to 5 mm) to determine how contact and gap influences lesion depth. The objective was to verify that the solver correctly captured the decrease in electric field intensity with increasing distance from the electrodes as measured in the experiments. 

\subsubsection*{Case 2: Spatial electric field distribution} 
\noindent This case benchmarks the spatial distribution of electric field with the study by \citet{Yao2017}. The solver was tested by simulating the field distribution around electrodes under applied voltages of 1000 V and 250 V. The setup consists of four pairs of parallel steel electrodes, with the left side positive and the right side negative. The spacing configuration included a 2.5 mm gap between positive and negative electrodes and a 2 mm separation between electrodes of the same polarity. The objective was to validate how the electric field propagates spatially around the electrodes.

\begin{figure}[H]
    \centering
    \begin{subfigure}[b]{0.95\textwidth}
        \centering
        \includegraphics[width=\textwidth]{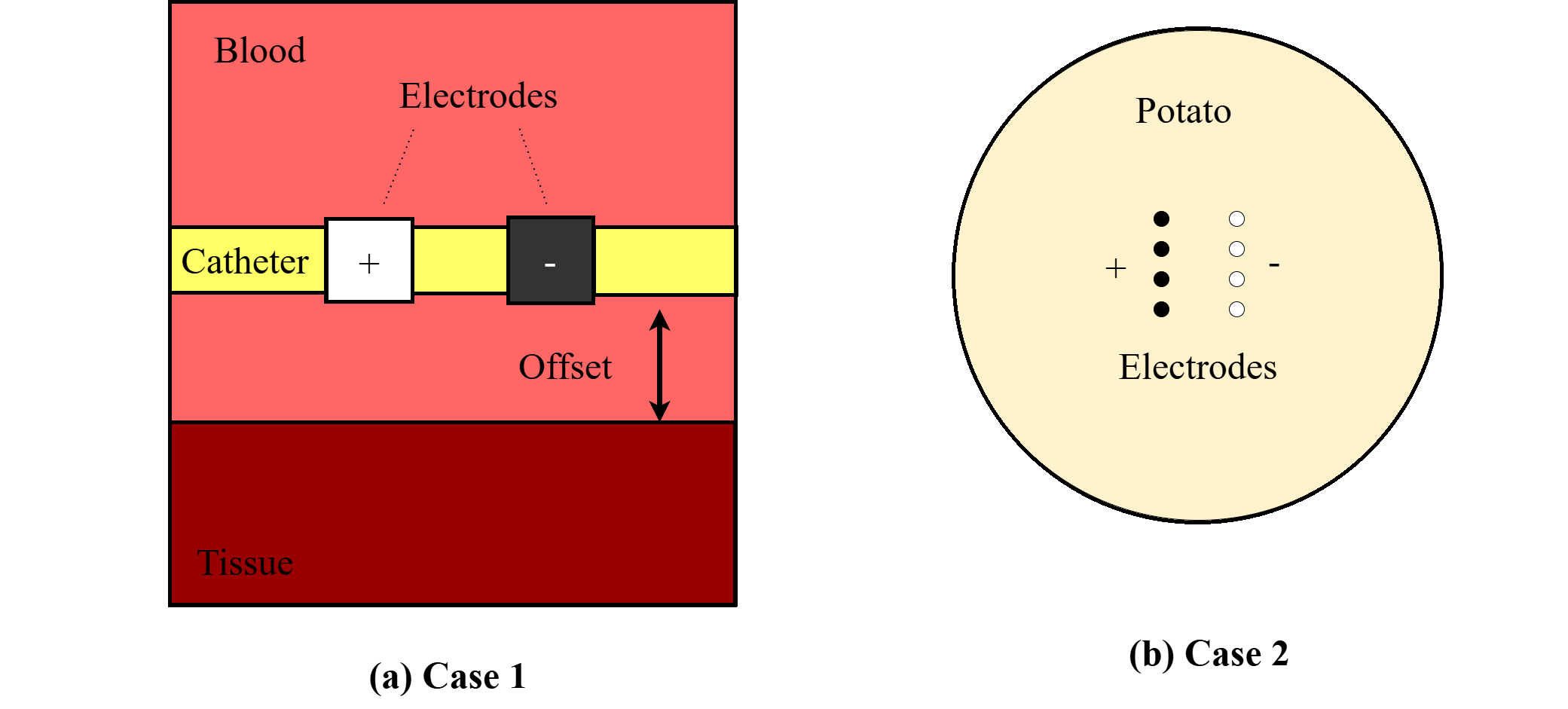}
        
    \end{subfigure}%
    \caption{Schematic views of (a) Case 1 from \citet{Meckes2022} and (b) Case 2 from \citet{Yao2017}}
\end{figure}

\subsubsection*{Case 3: Electric field and temperature rise validation}
\noindent This case evaluates the model's capability to predict both electric field distribution and temperature rise during PFA as found in  \citet{Arena2012}. A three-dimensional {\it in vitro} tumor platform was used to model irreversible electroporation and its effect on tissue heating. Electrodes with a diameter of 1.3 mm and a center-to-center spacing of 3.35 mm were inserted into a hydrogel-based tumor model in the experiment. Pulse characteristics included a duration of 100 $\mu s$, 80 pulses delivered at a repetition rate of one pulse per second. Applied voltages ranged from 150 V to 600 V, generating electric field strengths of up to 1800 V/cm. A fiber optic probe was placed between the electrodes, to monitor the temperature change.

\subsubsection*{Case 4: Lesion Formations with experimental study}
\noindent This case validates  the model's capability in predicting lesion formation under varying electrode-tissue distances. The experimental study by \citet{howard2022effects} analyzed lesion depth and width for different electrode offsets (0 mm, 2 mm, 4 mm) using a biphasic, bipolar PFA system. The model results are compared against these experimental measurements to assess its accuracy.

\begin{figure}[H]
    \centering
    \begin{subfigure}[b]{0.85\textwidth}
        \centering
        \includegraphics[width=\textwidth]{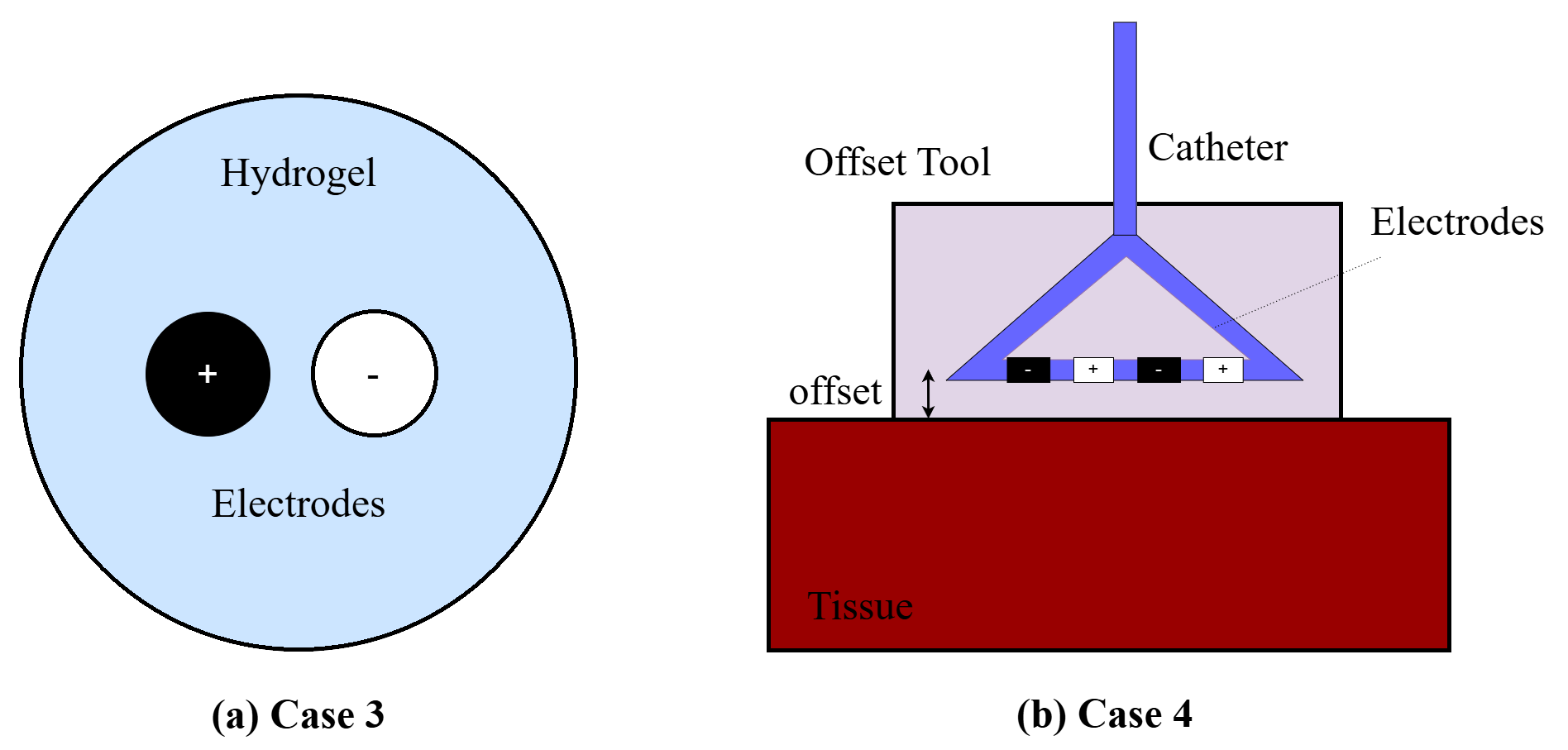}
    \end{subfigure}%
    \caption{Schematic views of (a) Case 3 from \citet{Arena2012} and (b) Case 4 from \citet{howard2022effects}}
\end{figure} 
\subsubsection*{Case 5: Ablation performance of different catheter configurations}

\begin{figure}[H]
    \centering
    
    \begin{subfigure}[b]{0.27\textwidth}
        \centering
        \includegraphics[width=\textwidth]{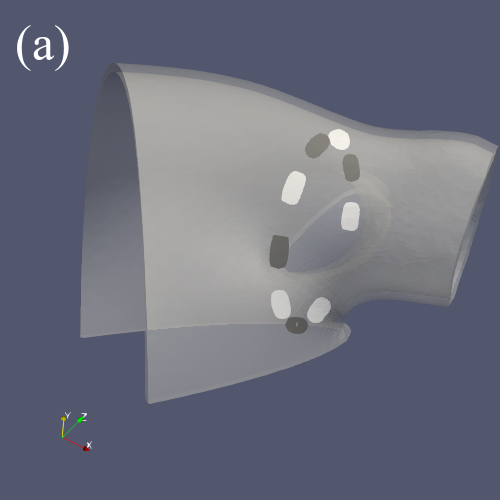}
    \end{subfigure}%
    \hspace{1pt}
    \begin{subfigure}[b]{0.27\textwidth}
        \centering
        \includegraphics[width=\textwidth]{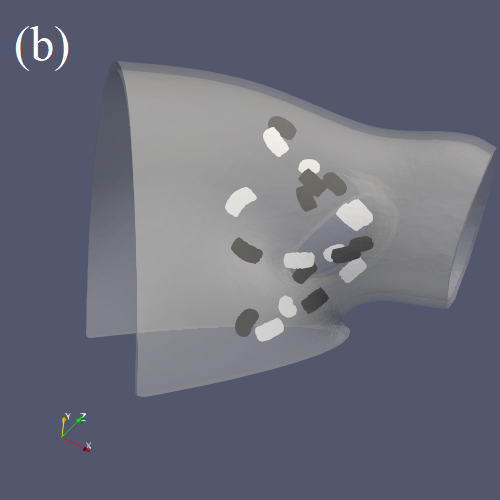}
    \end{subfigure}%
    \hspace{1pt}
    \begin{subfigure}[b]{0.27\textwidth}
        \centering
        \includegraphics[width=\textwidth]{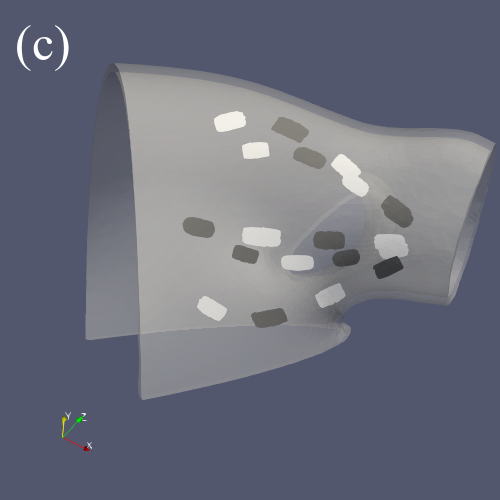}
    \end{subfigure}%
    \vspace{3pt}

    \begin{subfigure}[b]{0.27\textwidth}
        \centering
        \includegraphics[width=\textwidth]{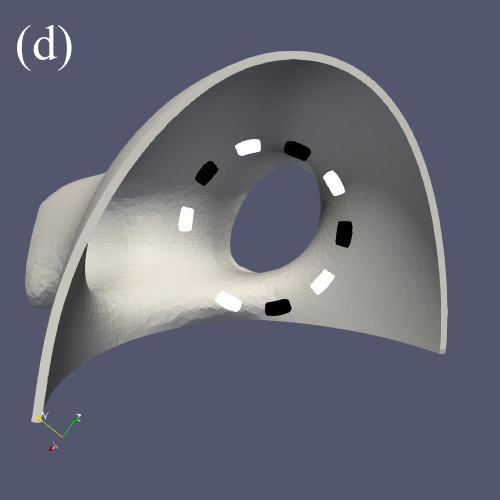}
    \end{subfigure}%
    \hspace{1pt}
    \begin{subfigure}[b]{0.27\textwidth}
        \centering
        \includegraphics[width=\textwidth]{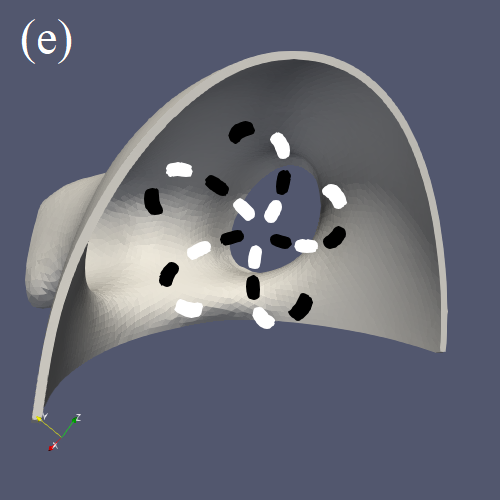}
    \end{subfigure}%
    \hspace{1pt}
    \begin{subfigure}[b]{0.27\textwidth}
        \centering
        \includegraphics[width=\textwidth]{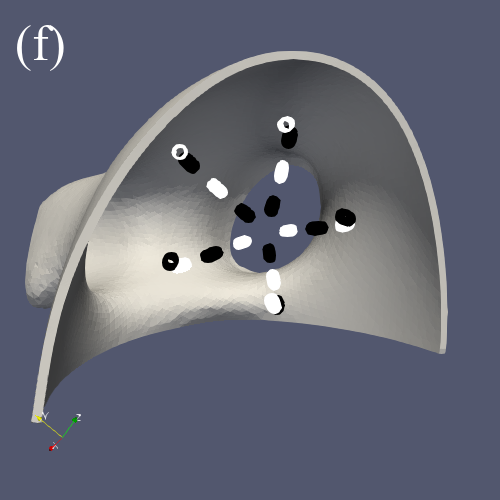}
    \end{subfigure}%
    \vspace{1pt}
    
    \caption{Catheter placements within the LSPV geometry. (a–c) Side views and (d–f) inner views of (a,d) circular, (b,e) flower, and (c,f) basket configurations} 
    \label{posit}
\end{figure}

\noindent After validating and verifying the model, we use it to assess the ablation performance of various catheter designs. The geometry of LSPV remains constant across all simulations, while different catheter configurations—including circular, flower, and basket types—are tested under a wide range of applied voltages.  In total, more than 100 simulations were performed. The positioning of each catheter type, shown in Fig. \ref{posit}, was designed to be as close as possible to the inner surface, mimicking optimal clinical deployment for effective energy delivery.
\noindent The circular catheter is positioned around the LSPV ostium, maintaining contact at the entrance. The flower catheter extends slightly inside and more outside the vein, offering broader surface coverage. The basket catheter penetrates fully into the LSPV, providing deeper contact through its radial splines.\noindent For the catheters, the voltage delivery pattern follows an alternating polarity configuration for the electrodes (positive in white, negative in black), as shown in Figure \ref{posit}.

As will be shown and discussed in Section \ref{rot-sec} one of the primary challenges during ablation is the uneven electric field distribution caused by anatomical irregularities and catheter design. For the circular catheter, the LSPV geometry may not fully align with the catheter shape, leading to regions with insufficient field coverage when a single position is used. Similarly, the flower and basket configurations inherently create ~72° gaps between branches, resulting in potential under-ablated areas. To address this issue, a two-step rotation strategy is adopted clinically for all catheter types to improve field overlap and produce more uniform and continuous lesions. For the circular catheter, an 18° rotation between applications helps compensate for geometric asymmetry. For the flower and basket catheters, a 36° rotation repositions the branches to fill the initial field gaps. The two-step approach was also simulated here. 
%The individual and combined electric field distributions, illustrated in Fig. \ref{rots}, demonstrate enhanced ablation uniformity and efficacy with this method.

To compare catheter efficacy and energy delivery characteristics, we defined six quantitative metrics: (1) the total volume of ablated tissue, calculated as the volume of tissue exposed to electric fields above 26.8~V/mm; (2) the energy delivery ratio, defined as the ratio of energy delivered to the target tissue versus the total applied energy, representing the efficiency of power deposition; (3) the normalized ablation power, defined as the ratio between lesion volume and the catheter’s surface area, to assess energy concentration efficiency; (4) the target tissue ablation percentage, which quantifies the percentage of a predefined anatomical target volume that is successfully ablated; (5) transmurality, defined as the percentage of the PV wall thickness that is ablated, indicating whether lesions span the full thickness of the target tissue; and
(6) Average lesion depth, defined as the mean distance from the tissue surface to the deepest point where the electric field exceeds the ablation threshold, providing a measure of how deeply the energy penetrates into the target tissue.

\color{black}

\section{Results}

\noindent Results for the validation cases of 1 to 4 show good agreement with the benchmark simulation and experiments (see Appendix). 
Here, we focus on the simulation results for different catheter configurations and applied voltages, which includes the electric field distributions, lesion characteristics, energy delivery performance and their comparison across the various setups.

\label{results}

%\subsection{Simulation results of different catheter configurations}
%This section presents the results demonstrating the effect of varying voltage and catheter configuration on ablated area characteristics, along with a comparative analysis.

\subsection{Comparison of electric field distribution between catheters in a single-step process}
\begin{figure}[H]

        \centering
    \begin{subfigure}[b]{0.9\textwidth}
        \centering
        \includegraphics[width=\textwidth]{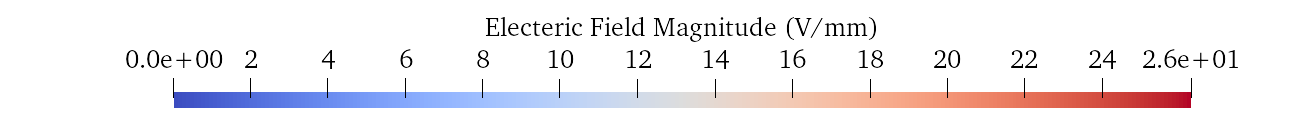}
    \end{subfigure}%
    
    \centering
    \begin{subfigure}[b]{0.48\textwidth}
        \centering
        \includegraphics[width=\textwidth]{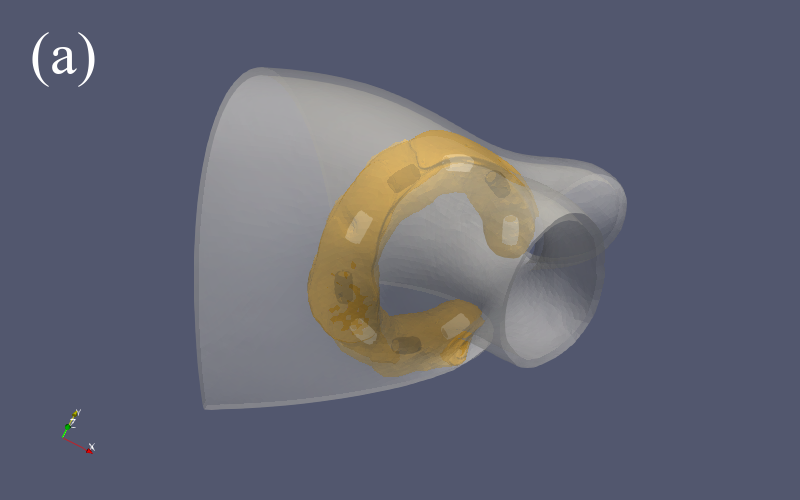}
    \end{subfigure}%
    \vspace{1pt}
    \begin{subfigure}[b]{0.48\textwidth}
        \centering
        \includegraphics[width=\textwidth]{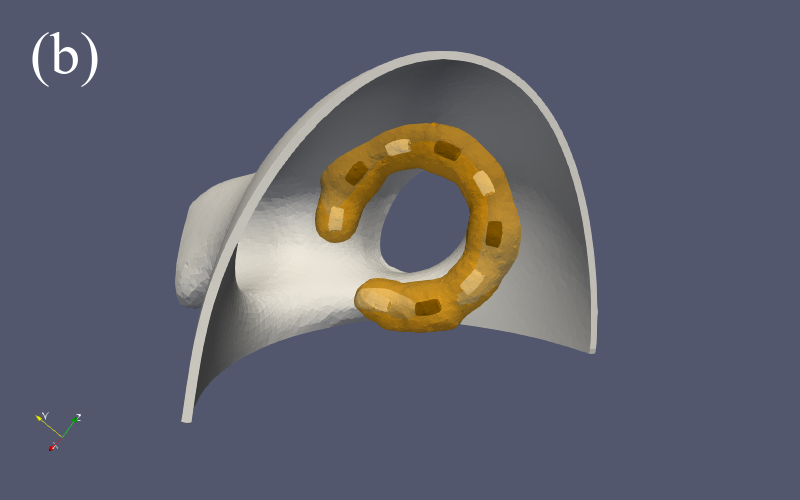}
    \end{subfigure}%
    \hspace{1pt}
    \begin{subfigure}[b]{0.48\textwidth}
        \centering
        \includegraphics[width=\textwidth]{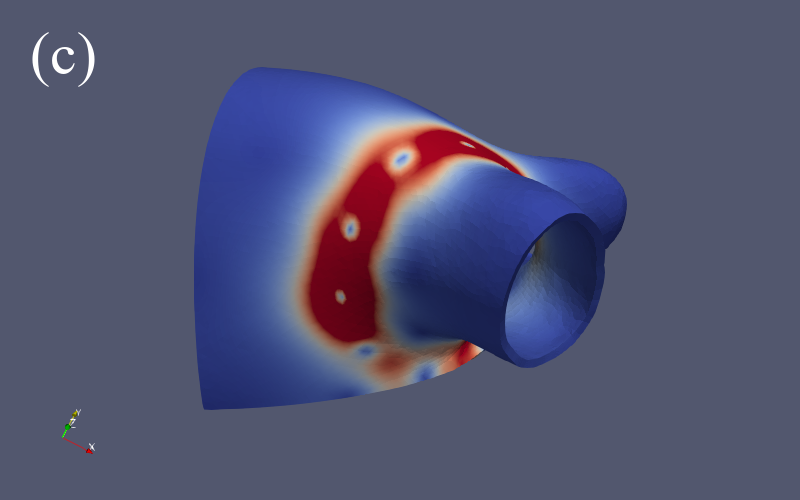}
    \end{subfigure}%
    \vspace{1pt}
    \begin{subfigure}[b]{0.48\textwidth}
        \centering
        \includegraphics[width=\textwidth]{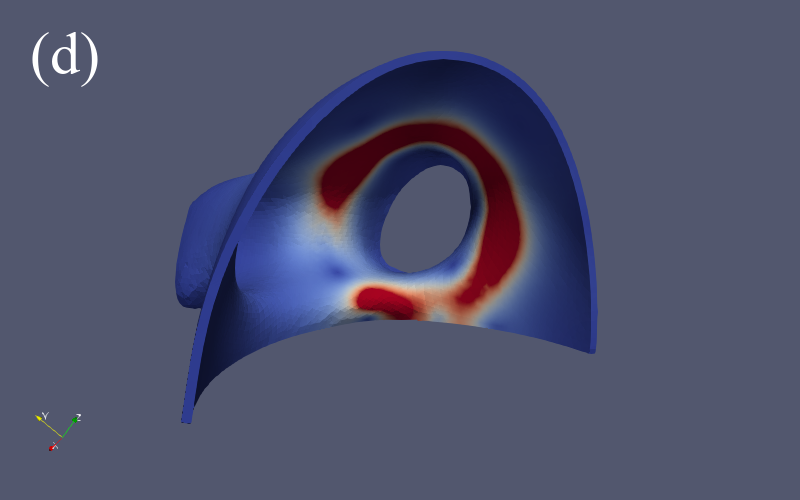}
    \end{subfigure}%
    \caption{Positioning of the circular catheter and the resulting electric fields in the (a),(b) blood and (c),(d) tissue during PFA for applied voltage of 500 V. Yellow regions in (a) and (b) indicate areas in the blood domain where the electric field exceeds the threshold. Red regions in (c) and (d) denote ablated tissue (where the electric field magnitude exceeds the threshold)
}
    \label{fig:circular}
\end{figure}

\noindent{\it Circular Configuration:} Figure \ref{fig:circular} shows the electric field distribution in both the blood and tissue domains during one stage of pulsed field ablation with the circular catheter configuration. As illustrated in the figure it generates a torus-shaped electric field around the electrodes. The resulting toroidal field promotes deeper, more localized lesion formation, ensuring a uniform circumferential pattern of ablation even at lower voltages.

\begin{figure}[H]

        \centering
    \begin{subfigure}[b]{0.9\textwidth}
        \centering
        \includegraphics[width=\textwidth]{Res-Figs/leg.png}
    \end{subfigure}%
    
    \centering
    \begin{subfigure}[b]{0.48\textwidth}
        \centering
        \includegraphics[width=\textwidth]{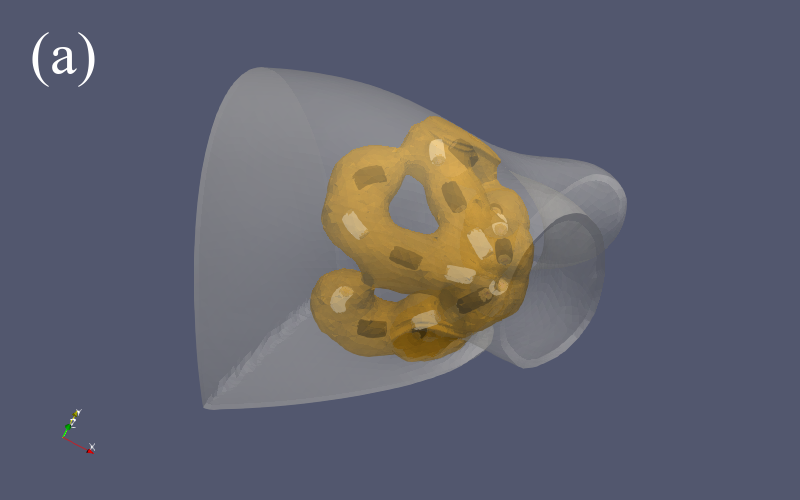}
    \end{subfigure}%
    \vspace{1pt}
    \begin{subfigure}[b]{0.48\textwidth}
        \centering
        \includegraphics[width=\textwidth]{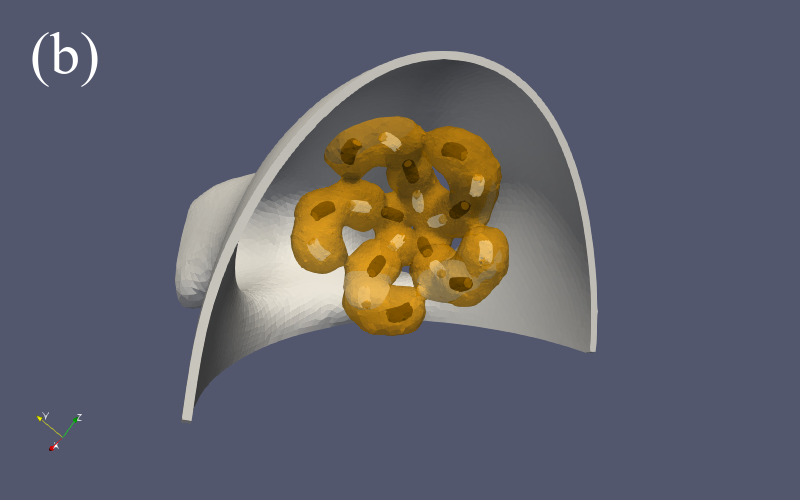}
    \end{subfigure}%
    \hspace{1pt}
    \begin{subfigure}[b]{0.48\textwidth}
        \centering
        \includegraphics[width=\textwidth]{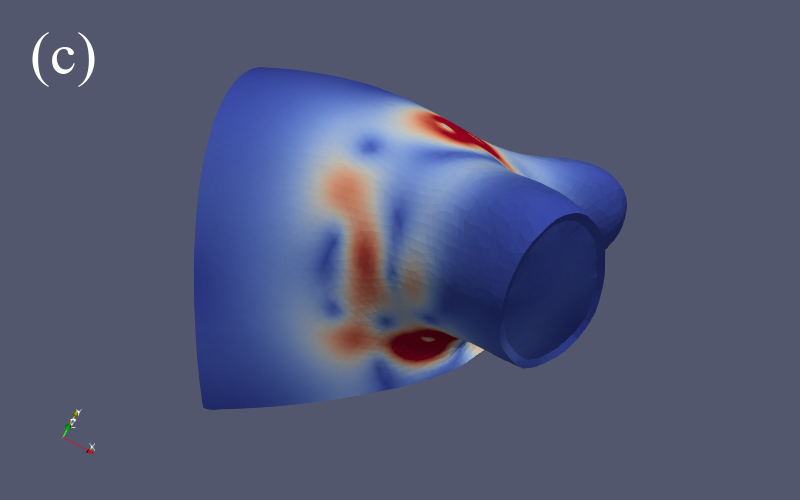}
    \end{subfigure}%
    \vspace{1pt}
    \begin{subfigure}[b]{0.48\textwidth}
        \centering
        \includegraphics[width=\textwidth]{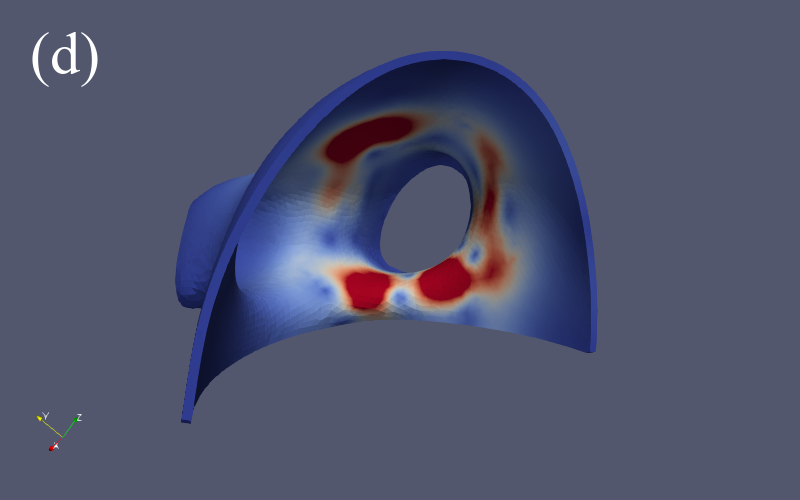}
    \end{subfigure}%
    \caption{Positioning of the flower configuration of the catheter and the resulting electric fields in the (a),(b) blood and (c),(d) tissue during PFA for applied voltage of 500 V (Color coding is consistent with Figure \ref{fig:circular})}
    \label{fig:flower}
\end{figure}

\noindent{\it Flower Configuration}: Figure~\ref{fig:flower} shows the flower catheter configuration and the resulting electric field distribution. In contrast to the circular catheter, the flower configuration generates a star-shaped pattern of electric fields due to its radially distributed electrodes. This pattern consists of multiple overlapping electric fields that extend circumferentially inside the PV, offering enhanced coverage and more uniform tissue contact. However, the angular offset of the petals leads under-ablation in certain regions, resulting in gaps in lesion continuity.

\begin{figure}[H]

        \centering
    \begin{subfigure}[b]{0.9\textwidth}
        \centering
        \includegraphics[width=\textwidth]{Res-Figs/leg.png}
    \end{subfigure}%
    
    \centering
    \begin{subfigure}[b]{0.48\textwidth}
        \centering
        \includegraphics[width=\textwidth]{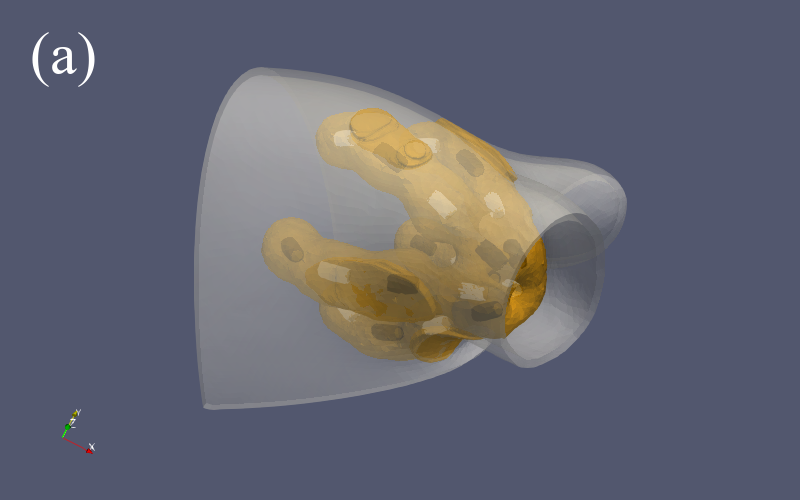}
    \end{subfigure}%
    \vspace{1pt}
    \begin{subfigure}[b]{0.48\textwidth}
        \centering
        \includegraphics[width=\textwidth]{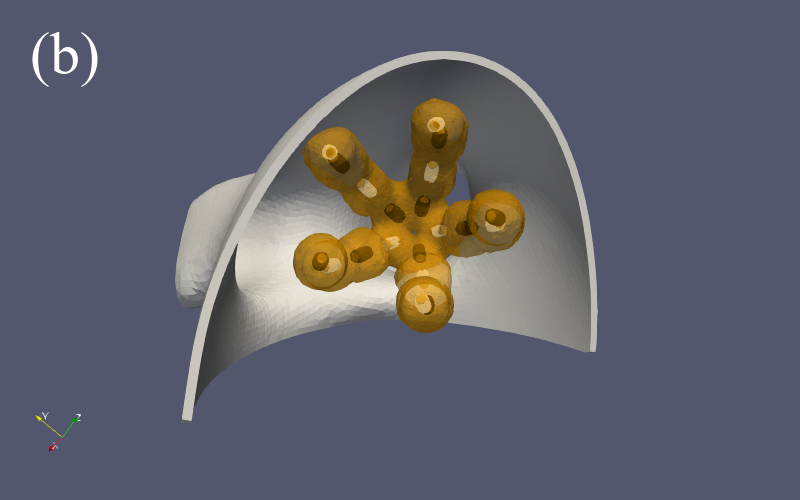}
    \end{subfigure}%
    \hspace{1pt}
    \begin{subfigure}[b]{0.48\textwidth}
        \centering
        \includegraphics[width=\textwidth]{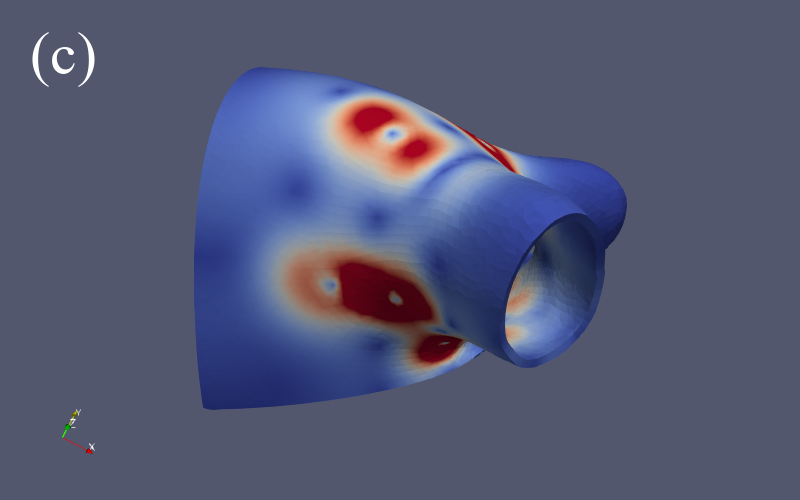}
    \end{subfigure}%
    \vspace{1pt}
    \begin{subfigure}[b]{0.48\textwidth}
        \centering
        \includegraphics[width=\textwidth]{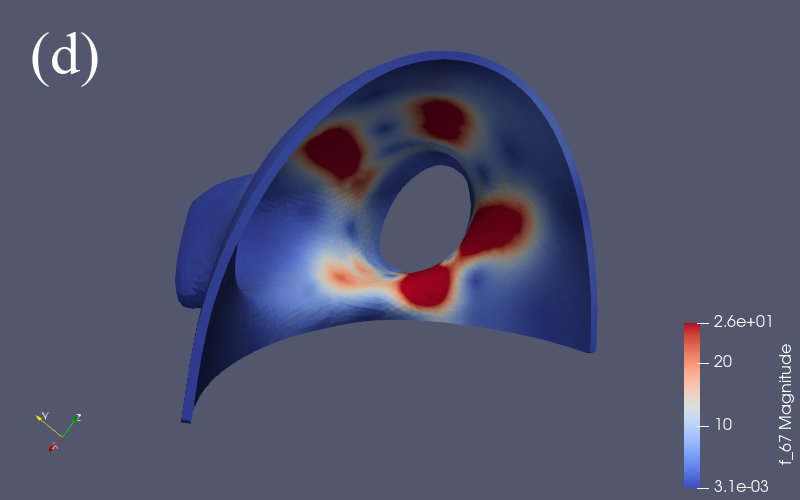}
    \end{subfigure}%
    \caption{Positioning of the basket configuration of the catheter and the resulting electric fields in the (a),(b) blood and (c),(d) tissue during PFA for applied voltage of 500 V}
    \label{fig:basket}
\end{figure}

\noindent {\it Basket Configuration}: \noindent Figure~\ref{fig:basket} shows the basket catheter configuration and the resulting electric field distribution during PFA. Similar to the flower catheter, it produces a radial pattern of electric fields. The basket’s deeper placement inside the vein leads to more localized and deeper tissue ablation. However, despite the increased ablation depth, the field distribution remains discontinuous due to the wide spacing between branches, resulting in persistent gaps that leave regions untreated.

%LC : Perhaps we can combine Fig. 7 - 9 into 1 figure
%AB : I tried Combining the figures but the result in terms of size and details was not good.

\subsection{Effects of Two-Step Ablation Process and Catheter Rotation}
\label{rot-sec}
\begin{figure}[H]
    \centering
    \begin{subfigure}[b]{0.8\textwidth}
        \centering
        \includegraphics[width=\textwidth]{Res-Figs/leg.png}
    \end{subfigure}%
    \centering
    \vspace{3pt}
        \centering
    \vspace{3pt}
        \centering
    \begin{subfigure}[b]{0.9\textwidth}
        \centering
        \includegraphics[width=\textwidth]{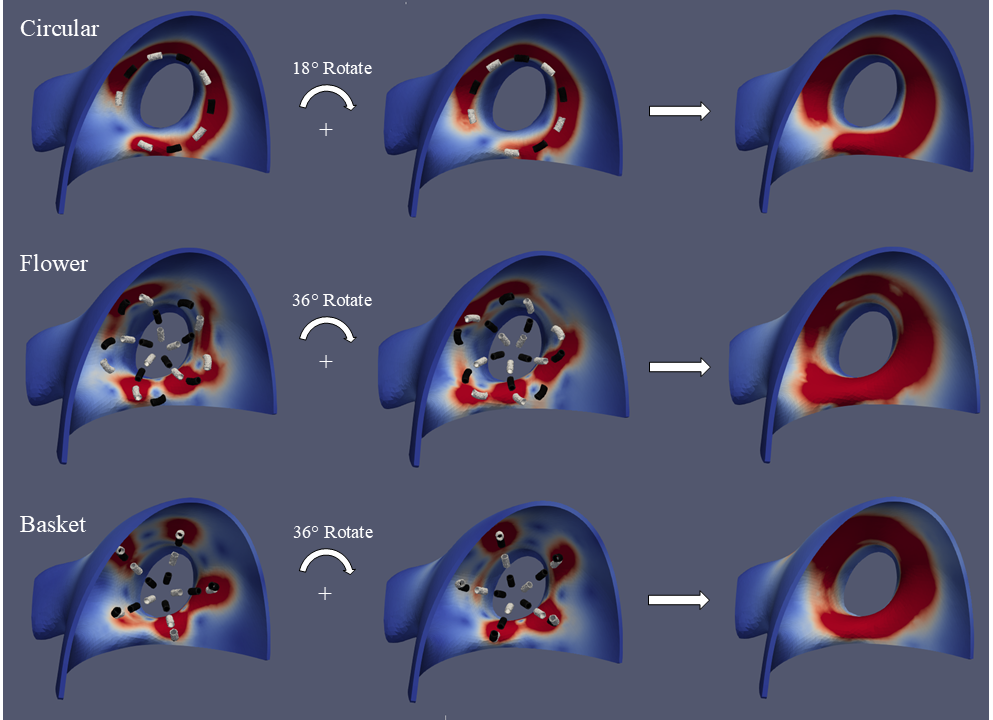}
    \end{subfigure}%
    
    \centering

    \caption{Two-step ablation strategy and electric field distribution for circular, flower and basket catheter configurations for applied voltage of 500 V}
    \label{rots}    
\end{figure}

\noindent Figure\ref{rots} shows the simulation results from the two-step ablation approach applied to the circular, flower, and basket catheter configurations. In this approach, the catheter is rotated—by 18 degrees for the circular and 36 degrees for the flower and basket designs—before a second ablation is performed. This process significantly improves electric field uniformity at the pulmonary vein ostium, effectively covering previously under-ablated regions. The combined result yields greater lesion volume and depth across all configurations. All subsequent results and analyses in the following sections are based on this two-stage ablation approach and reflect the total cumulative effect of the two stages.

\subsection{Effects of applied voltage Magnitude}

\begin{figure}[H]
    \centering
    \begin{subfigure}[b]{0.8\textwidth}
        \centering
        \includegraphics[width=\textwidth]{Res-Figs/leg.png}
    \end{subfigure}%
    \centering
    \vspace{3pt}
        \centering
    \begin{subfigure}[b]{0.8\textwidth}
        \centering
        \includegraphics[width=\textwidth]{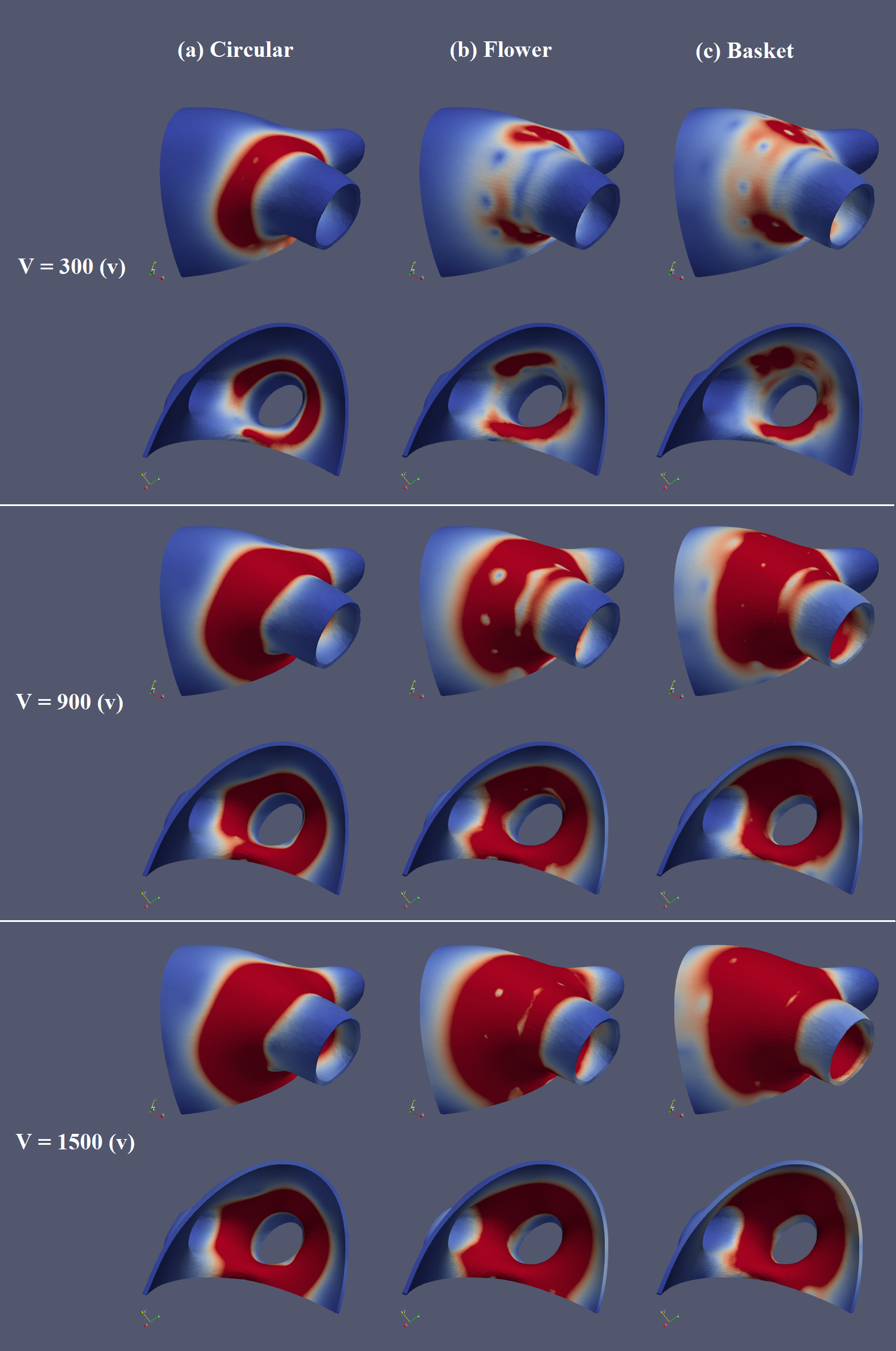}
    \end{subfigure}%
    \centering
    \caption{Electric field distribution magnitude for all catheters at increasing voltages: 300, 900, and 1500 V}  
    
    \label{voltages}
\end{figure}

\noindent Figure~\ref{voltages} shows the electric field distribution for the circular, flower, and basket catheter configurations at different voltages. For all configurations, higher voltages result in larger regions exceeding the ablation threshold, leading to more extensive and deeper lesion formation. At 300 V, electric field coverage is limited and fragmented. By 900 V, continuous lesion formation becomes evident, particularly in the basket configuration. At 1500 V, all three catheters produce strong, uniform electric fields around the PV ostium, indicating effective ablation across the targeted tissue.  

\subsection{Lesion Transmurality}
\begin{figure}[H]
    \centering
    \begin{subfigure}[b]{0.8\textwidth}
        \centering
        \includegraphics[width=\textwidth]{Res-Figs/leg.png}
    \end{subfigure}%
    \centering
    \vspace{3pt}
        \centering
        \includegraphics[width=1.0\textwidth]{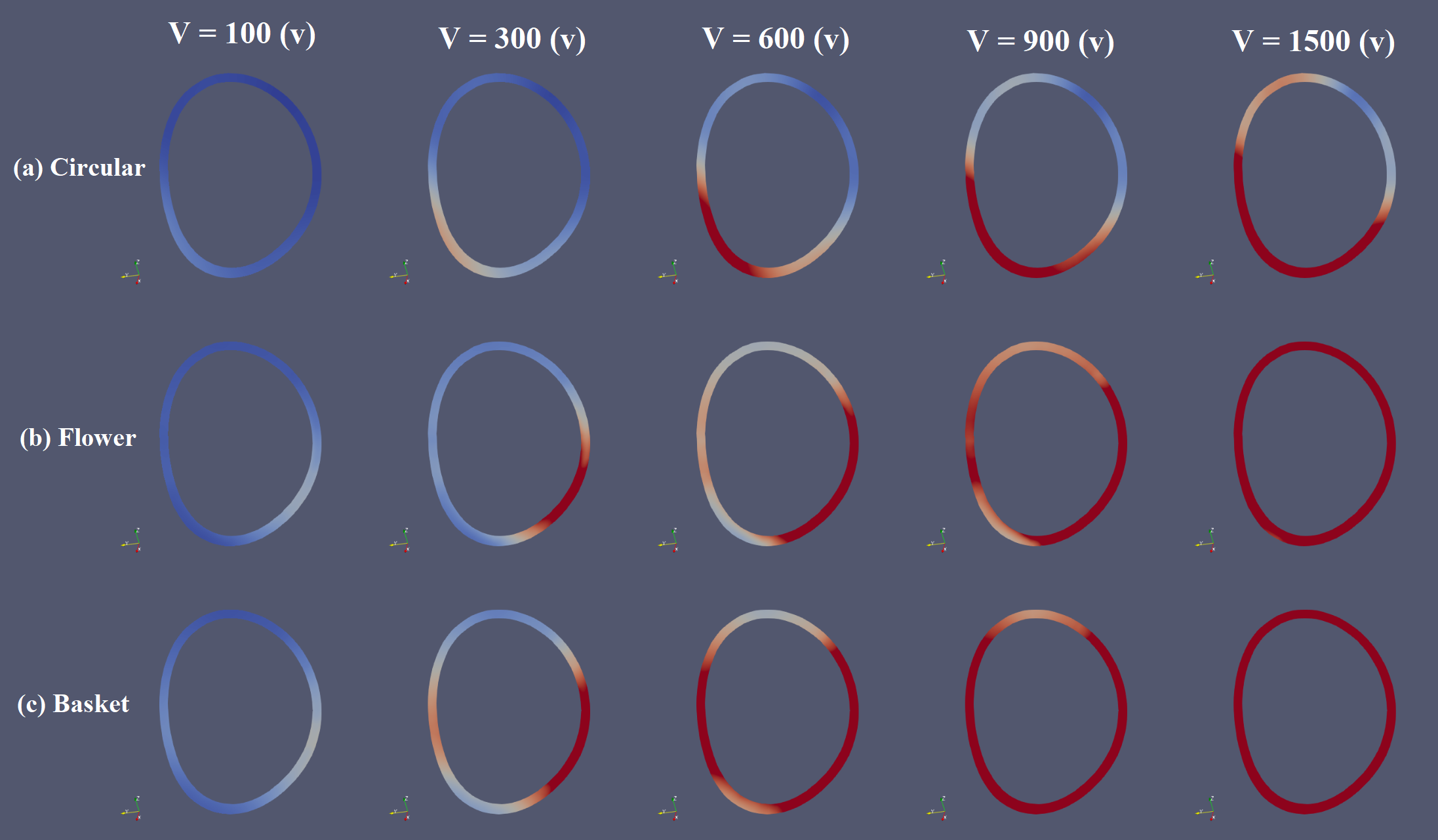}
    \caption{Cross-sectional electric field magnitude at the mid-plane of the target tissue for circular (a), flower (b), and basket (c) catheter configurations at voltages of 100, 300, 600, 900, and 1500 V}
        \label{trans}
\end{figure}

\noindent Figure~\ref{trans} shows cross-sectional views at the mid-plane of the target tissue for the circular, flower, and basket catheter configurations across a range of voltages (100 V to 1500 V). At low voltages (100–300 V), the electric field intensity is insufficient to penetrate the full thickness of the tissue, resulting in non-transmural lesions. As the applied voltage increases, the regions exceeding the ablation threshold (26.8 V/mm) extend progressively through the tissue wall. At 900 V and especially 1500 V, all three configurations achieve near-complete transmurality in middle section, with the basket catheter showing the most consistent through-wall field coverage.
\noindent

\subsection{Comparison of Efficiency and Performance of the catheters}
\begin{figure}[H]

    \centering
    \begin{subfigure}[b]{0.48\textwidth}
        \centering
        \includegraphics[width=\textwidth]{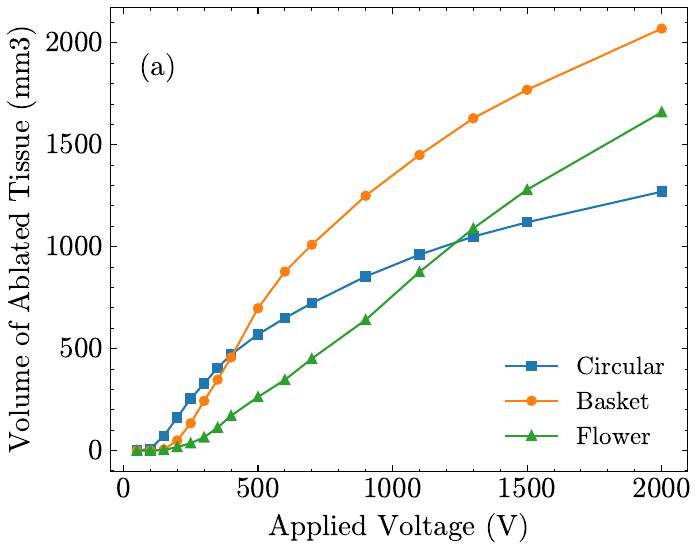}
    \end{subfigure}%
    \vspace{1pt}
    \begin{subfigure}[b]{0.48\textwidth}
        \centering
        \includegraphics[width=\textwidth]{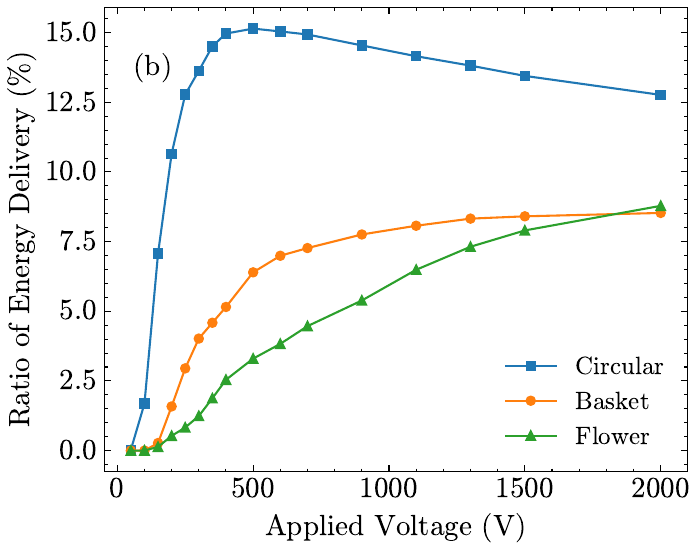}
    \end{subfigure}%
    \hspace{1pt}
    \begin{subfigure}[b]{0.48\textwidth}
        \centering
        \includegraphics[width=\textwidth]{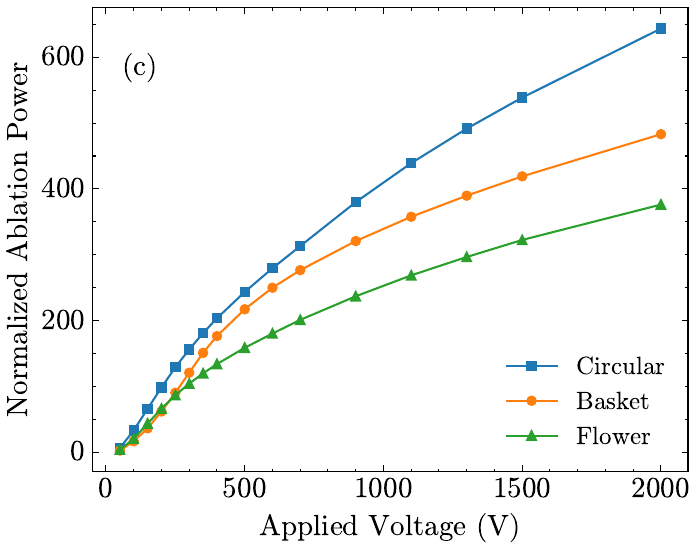}
    \end{subfigure}%
    \vspace{1pt}
    \begin{subfigure}[b]{0.48\textwidth}
        \centering
        \includegraphics[width=\textwidth]{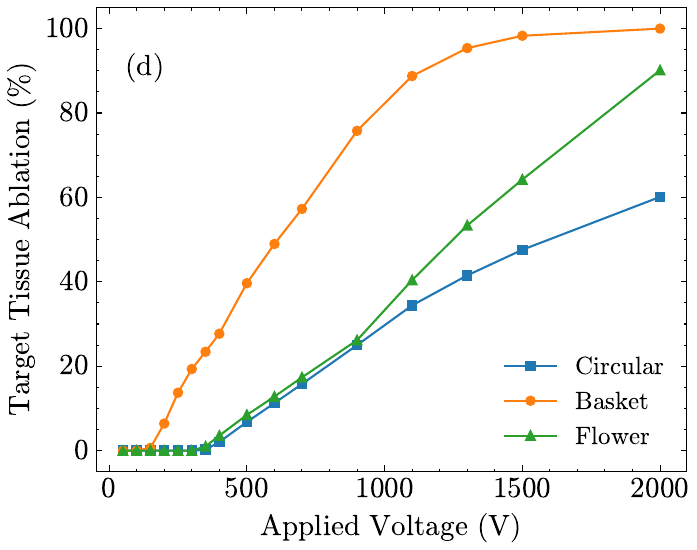}
    \end{subfigure}%
    \hspace{1pt}
    \begin{subfigure}[b]{0.48\textwidth}
        \centering
        \includegraphics[width=\textwidth]{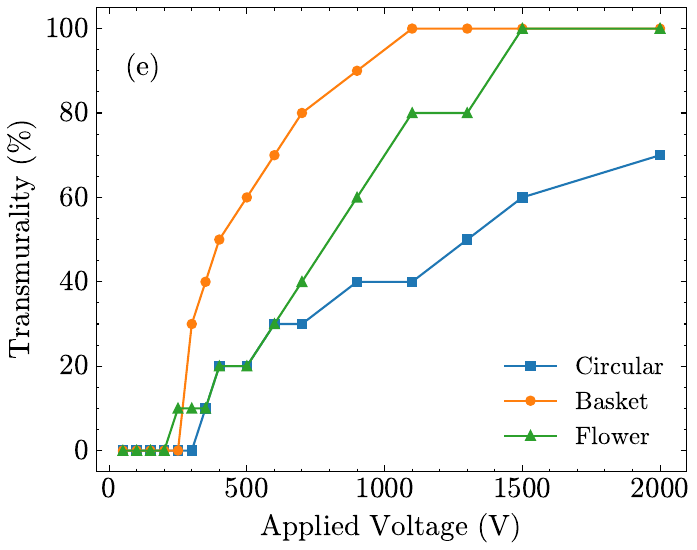}
    \end{subfigure}%
    \vspace{1pt}
    \begin{subfigure}[b]{0.48\textwidth}
        \centering
        \includegraphics[width=\textwidth]{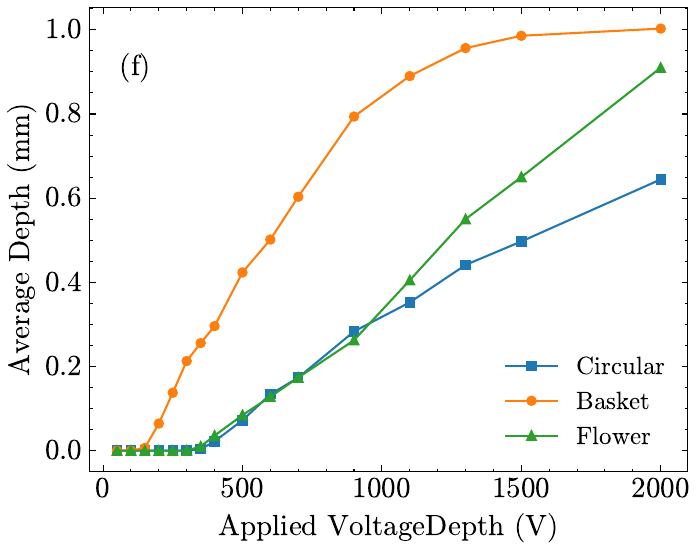}
    \end{subfigure}%
    \caption{ Comparison of catheters performance over a voltage range of 50–2000 V}
    \label{fig:metrics}
\end{figure}

\noindent Figure~\ref{fig:metrics} quantitatively summarizes the  performance and efficiency of the circular, flower, and basket catheter configurations across increasing applied voltages using six quantitative metrics.  Among these 3 catheters, the circular catheter has the largest normalized ablation power and energy delivery ratio, reaching approximately 15\% at 500 V.  The basket catheter, however, produces the largest overall lesion volumes, target tissue coverage, transmurality, and average lesion depth across all voltages, where full transmurality and complete target ablation are achieved at 1500 V. In contrast,  only about 60\% and 90\% of the targeted tissue is ablated at 1500 V by the circular and flower catheter, respectively. 
%LC: You mention about heat conduction in your governing equation but did not say anything about it in the results. If the temperature increase is small, it may be worthwhile to mention it here that the rise in temperature is about a few degree etc.?
%AB : I didn't include the temperature with 3 main catheter because the voltage delivary time is usually too short and there is not significant temperature rise. based on my hand calculation it is even less than 0.01 C. 

\color{black}

\section{Discussion}
\label{discussion}
\noindent We have developed a computational framework based on an open-source FE library to simulate PFA with different catheters in a realistic LA geometry.
To establish the credibility of the computational framework, an extensive validation process was conducted using both experimental studies and previously published numerical simulations. Most of the benchmark cases employed commercial solvers, such as COMSOL Multiphysics, facilitating a direct and consistent comparison across different modeling scenarios with our framework. The validation covered key aspects of PFA, including electric field distribution, temperature rise, and lesion dimensions. Across all validation cases, the present model showed strong agreement with reference results, yielding an average relative error of less than 7\% and a maximum deviation of approximately 15\%. These results confirm the computational framework capability to reliably reproduce relevant biophysical phenomena.

Our results show that the catheter geometry and the layout of its electrodes  can significantly influence distinct energy delivery characteristics. Circular catheters typically employ electrodes arranged in a loop, offering focused, symmetrical energy delivery. In contrast, flower-shaped catheters tend to distribute energy more diffusely, which may limit their ability to concentrate power on a localized area. Basket catheters, forming an expanded structure going deeper into the vein, provide broader coverage but often at the cost of reduced focal intensity due to increased exposure to the surrounding blood pool. 

Between these 3 catheters, the circular catheter demonstrated high energy delivery efficiency and produced compact lesions at relatively low voltages, reflecting its strength in focused power deposition and efficiency per unit surface area. The circular catheter has the best performance in terms of normalized ablation power and energy delivery ratio. This efficiency, however, does not translate into ablative effectiveness, as the circular catheter fails to achieve high levels of target tissue ablation and transmurality, reaching only around 60\% at 1500 V. This is largely due to its positioning around the pulmonary vein, which limits its ability to penetrate deeply and uniformly into the tissue. In contrast, the basket catheter produced the largest overall lesion volumes and, while its energy deposition was more widespread and less focused, it resulted in significantly greater target tissue coverage, transmurality, and average lesion depth.  The flower catheter, although showing performance closer to the basket catheter in some metrics—particularly transmurality and target coverage—still lags in both lesion volume and depth. Importantly, the fact that both the basket and flower catheters extend deeper into the vein gives them the capability to achieve faster and more complete ablation of the entire target region. Overall, the basket catheter offers the best balance between lesion effectiveness and coverage, making it the most effective option among the three designs for this anatomical setup.
 
These findings also align with those reported by \citet{belalcazar2024comparison} in their computational study comparing various PFA catheter designs, including circular and penta-spline (basket-equivalent) geometries. They observed that circular catheters, owing to their tightly spaced, bipolar electrode arrangement, achieved significantly higher energy delivery efficiency whereas the penta spline basket design demonstrated largest lesions but with less focused energy deposition. 

%LC : How is our study different from them? Need to mention here
% I can't Understand if you mean form methods point of view or in terms of findings 

Beyond comparison of different catheters' performance, this computational framework can potentially be used for clinical decision support and treatment planning. One of its impactful applications is in the optimization of catheter positioning with respect to the unique patient-specific pulmonary vein (PV) geometry. By simulating various catheter placements and orientations, the model can identify configurations that maximize lesion completeness and transmurality while minimizing energy waste and untreated regions. Moreover, it enables the selection of the most suitable catheter design (circular, basket, or flower) based on the anatomical shape and size of the PV ostium. For example, narrower or more curved ostia may benefit from circular configurations, while wider, deeper veins may require basket or flower catheters for effective lesion formation. In future applications, this model can be extended with the integration of machine learning algorithms to support real-time intraoperative planning or pre-procedural mapping, helping clinicians tailor the ablation strategy to the patient’s specific anatomy for improved safety and efficacy.

There are several limitations associated with this study. First, we assumed that the myocardial tissues exhibit homogeneous properties and fixed electroporation thresholds. We also did not consider other factors such as anisotropic conductivity, hydration variability, and imperfect electrode-tissue contact, gas bubble dynamics that may be relevant.
Second, we assumed quasi-static energy delivery and do not consider PFA waveforms, which are proprietary to the manufacturers.
Third, we omit the dynamic interplay of blood flow, tissue deformation, and motion that can influence lesion formation in vivo. 
%Last, Catheter placement, although designed to resemble clinical positioning, is arbitrarily defined and may not fully capture real-world variability. Prior studies, including \citet{belalcazar2024comparison}, emphasize the impact of waveform uncertainties, thermal effects, and gas bubble dynamics—factors not modeled here. 
Future work will aim to overcome these challenges by integrating electromechanical coupling, flow dynamics, and advanced tissue characterization to enhance clinical applicability and model fidelity.

\section{Conclusion}
\label{conclusion}

\noindent In summary, we developed a validated 3D computational modeling framework for simulating PFA in patient-specific LA anatomy. This study systematically investigated the effects of the catheter configuration and applied voltage on lesion morphology, volume, transmurality, and energy delivery efficiency. Three commonly used catheter designs—circular, flower, and basket—were analyzed across a broad voltage range to evaluate their performance in achieving effective ablation. Results showed that circular catheters provided the highest energy delivery efficiency and compact lesions at lower voltages, but were limited in depth and uniformity due to their geometry. Basket catheters, by contrast, produced the largest lesion volumes and achieved superior transmurality and target coverage, making them the most effective overall despite less focused energy deposition. Flower catheters demonstrated intermediate performance but required higher voltages to match the effectiveness of the other designs. The model's predictions were in strong agreement with experimental and computational benchmarks, underscoring its reliability. Beyond performance assessment, this framework offers a powerful platform for optimizing catheter selection and positioning based on individual PV anatomy, with potential applications in pre-procedural planning and real-time procedural guidance to enhance the safety and efficacy of PFA interventions.

\section*{Acknowledgments}
\noindent This work was supported by the MSU DFI grant. T.G. is partially supported by NSF CMMI grant No. 2323917.

\bibliography{cas-refs}
\bibliographystyle{unsrtnat}

\newpage
\section*{Appendix 1 - Validation Cases}

\subsection*{Case 1}
\label{valida}
\begin{figure}[H]
    \centering
    \begin{subfigure}[b]{0.5\textwidth}
        \includegraphics[width=\textwidth]{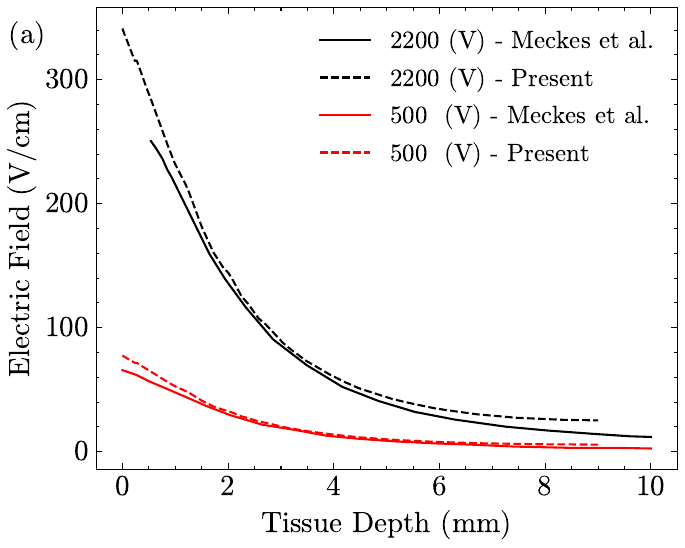}
    \end{subfigure}%
    \begin{subfigure}[b]{0.42\textwidth}
        \includegraphics[width=\textwidth]{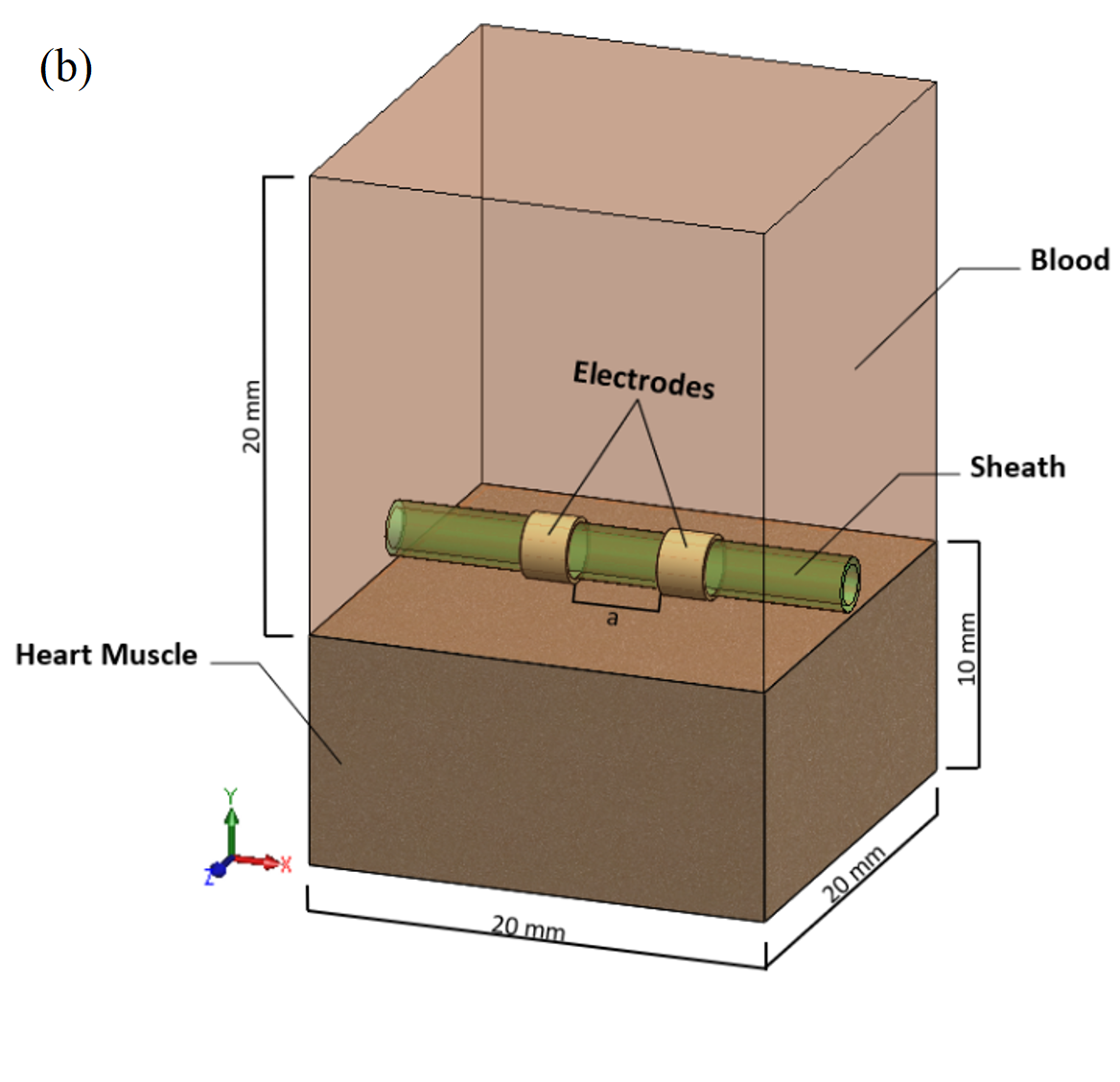}
    \end{subfigure}%
    \caption{Comparison of electric field magnitude from \citet{Meckes2022} data and present simulations for different applied voltages (left). The 3D model geometry shows the electrode placement used in \cite{Meckes2022} (right).} 
    \label{fig:val1}
\end{figure}

\noindent  Figure \ref{fig:val1} shows the variation electric field magnitude with tissue depth predicted by the model and that found in the benchmark case \cite{Meckes2022} for different voltages. There is a strong agreement between the model prediction and the benchmark results (maximum difference of 5\%), where both show an exponential decay of the electric field magnitude with depth.

\subsection*{Case 2}
\begin{figure}[H]
    \centering
    \begin{subfigure}[b]{0.95\textwidth}
        \centering
        \includegraphics[width=\textwidth]{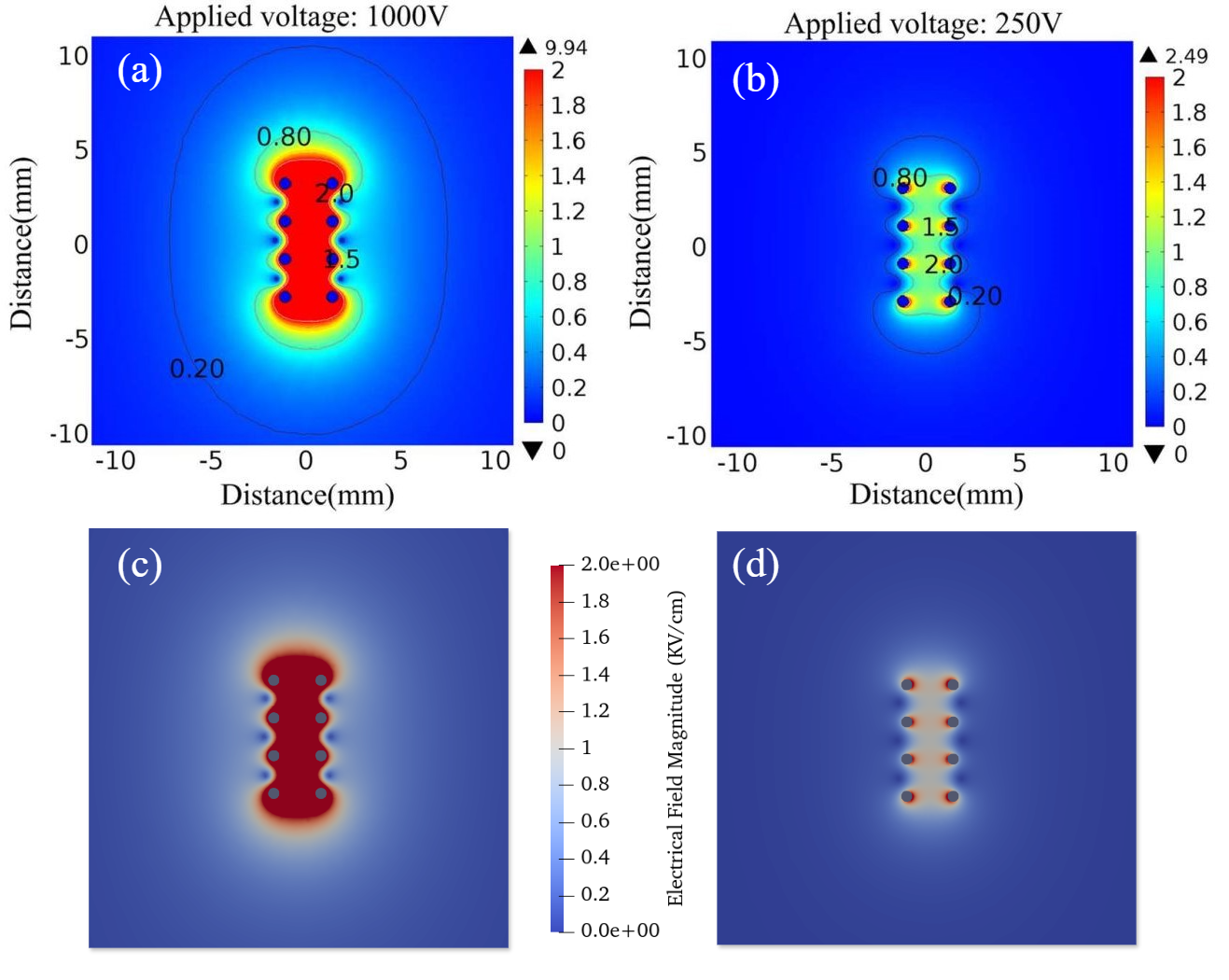}
    \end{subfigure}%
      
    \caption{Electric field distribution around electrodes from \citet{Yao2017} (a), (b) and corresponding present results (c), (d) for an applied voltages of 1000 V and 250 V. }
    \label{fig:val2}
\end{figure}

%\subsubsection{Second Case: Spatial Electric Field Distribution}
 \noindent Figure \ref{fig:val2} shows the model prediction of electric field distribution is close to that found in \citet{Yao2017}. For the two different applied voltages, the field magnitude decreases radially from the electrodes.

%\subsubsection{Third Case: Electric Field and Temperature Rise Validation}

\subsection*{Case 3}

\begin{figure}[H]
    \centering
    \begin{subfigure}[b]{0.4\textwidth}
        \centering
        \includegraphics[width=\textwidth]{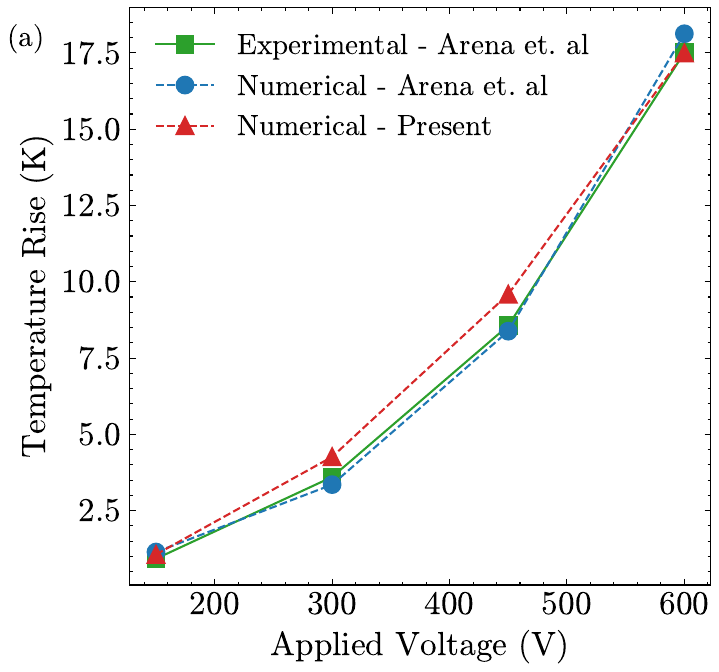}
    \end{subfigure}%
    \hfill
    \begin{subfigure}[b]{0.6\textwidth}
        \centering
        \includegraphics[width=\textwidth]{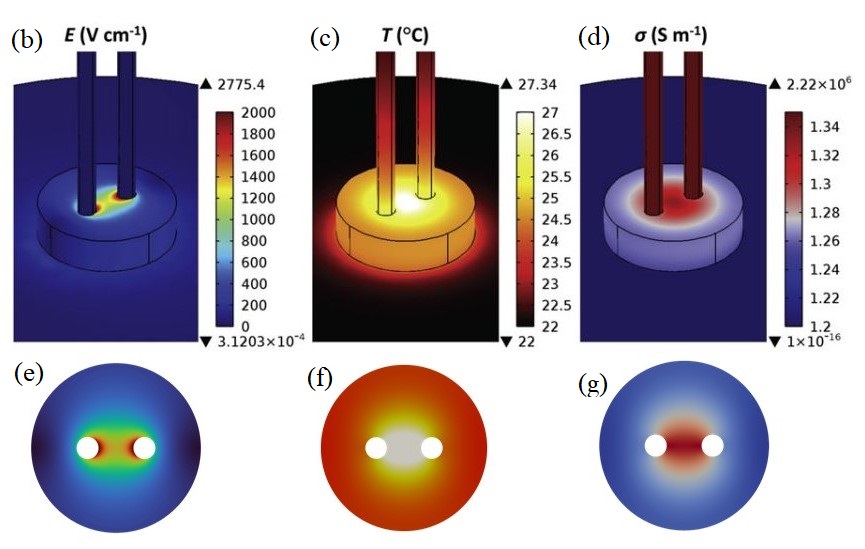}
    \end{subfigure}%
    \caption{(a) Comparison of temperature rise in \citet{Arena2012} and present cases, distribution of electric field, temperature, and conductivity near electrodes (b), (c), (d) for the 300 V case \cite{Arena2012} and present simulations (e), (f), (g).}
    \label{fig:val3}
\end{figure}

\noindent Figure~\ref{fig:val3} (a) shows the comparison of temperature rise between the model predictions and experimental measurements for different voltages with and average error of 12\% for temperature and 8\% for the electric field.  % of 150 V, 300 V, 450 V, and 600 V.  
%LC: Describe the variation. Also where is the temperature measured - in the middle? Error?
%Both the ablated area and the temperature rise have been validated against experimental data, confirming the solver’s accuracy in capturing the thermal and electric field effects in this case.
% AB: The probe place mentioned in the methods sction 

\subsection*{Case 4}

\begin{figure}[H]
    \centering
    \begin{subfigure}[b]{0.49\textwidth}
        \centering
        \includegraphics[width=\textwidth]{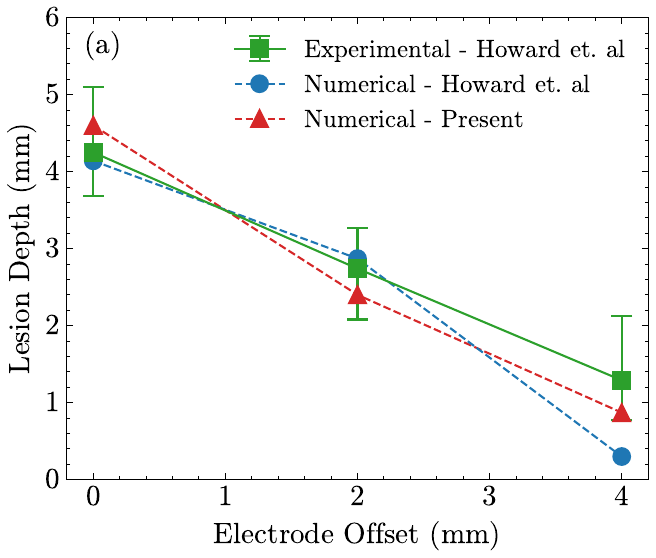}
    \end{subfigure}%
    \hfill
    \begin{subfigure}[b]{0.5\textwidth}
        \centering
        \includegraphics[width=\textwidth]{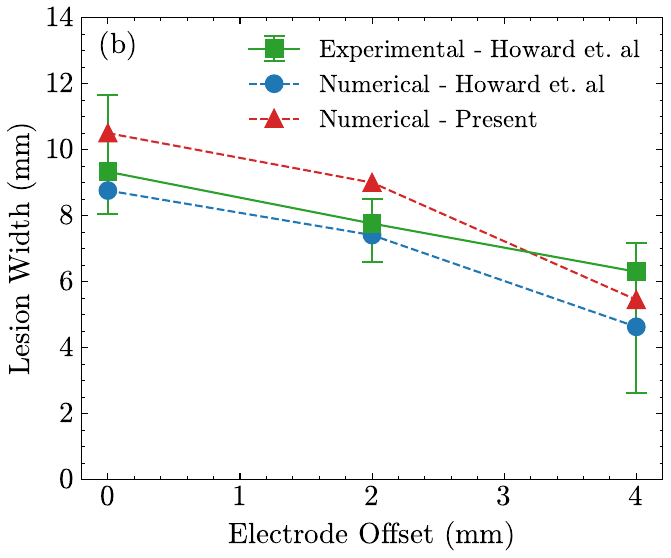}
    \end{subfigure}%
    \caption{Comparison of numerical and experimental (a) lesion depth and (b) lesion width results for different electrode offsets}
    \label{fig:val88}
\end{figure}

\noindent Figure \ref{fig:val88} shows the comparison of lesion depth and width predicted by the model with the experimental results.
Lesion depth and width decreases with increased electrode offset.

\begin{figure}[H]
    \centering
    \begin{subfigure}[b]{0.47\textwidth}
        \centering
        \includegraphics[width=\textwidth]{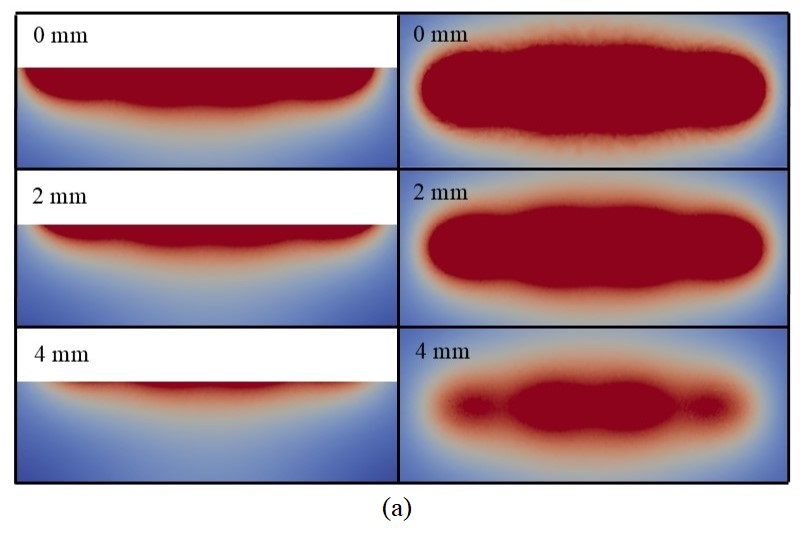}
    \end{subfigure}%
    \hfill
    \begin{subfigure}[b]{0.52\textwidth}
        \centering
        \includegraphics[width=\textwidth]{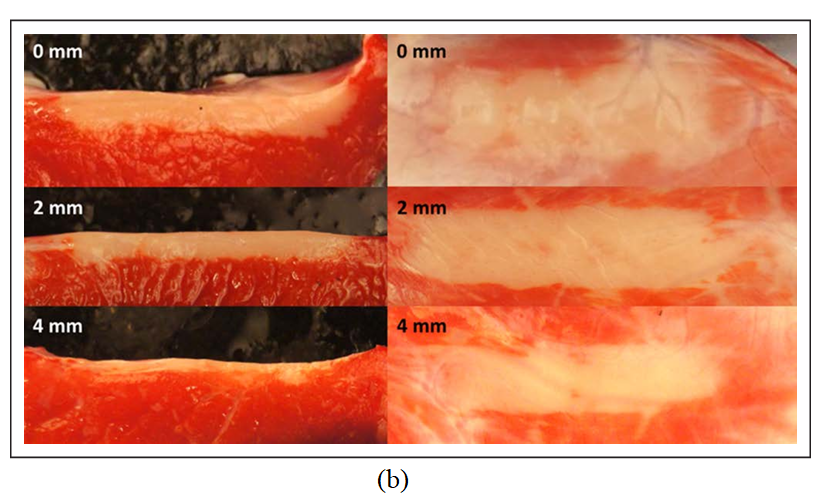}
    \end{subfigure}%
    \caption{(a) Present simulations and (b) experimental images from \citet{howard2022effects} (right) lesion depth for electrode offsets of 0 mm, 2 mm, and 4 mm}  
    \label{fig:val4}
\end{figure}

\pagebreak
\section*{Appendix 2 - Weak form and discretization}

In the weak formulations, \( V \) denotes the test and trial function space and \( v, w \in V \) are test functions. The solution fields \( \phi \) and \( T \) are also assumed to lie in \( V \). Let \( \Omega \subset \mathbb{R}^3 \) denote the computational domain representing the tissue and blood volumes. 

The electric potential $\phi$ satisfies:
\begin{equation}
\nabla \cdot (\sigma \nabla \phi) = 0
\end{equation}

where $\sigma(T) = \sigma_0 (1 + \alpha(T - T_b))$ is the temperature-dependent electrical conductivity.
In the weak form we find $\phi \in V$ such that

\begin{equation}
\int_\Omega \sigma \nabla \phi \cdot \nabla v \, d\Omega = 0 \quad \forall v \in V.
\end{equation}

The tissue temperature $T$ evolves according to:
\begin{equation}
\rho_t c_p \frac{\partial T}{\partial t} = \nabla \cdot (k_t \nabla T) + \sigma |\nabla \phi|^2 - \rho_b c_b \omega_b (T - T_b)
\end{equation}

In the weak form we find $T \in V$ such that
\begin{equation}
\int_\Omega \rho_t c_p \frac{\partial T}{\partial t} w \, d\Omega
+ \int_\Omega k_t \nabla T \cdot \nabla w \, d\Omega
+ \int_\Omega \rho_b c_b \omega_b (T - T_b) w \, d\Omega
= \int_\Omega \sigma |\nabla \phi|^2 w \, d\Omega \quad \forall w \in V.
\end{equation}

Using Backward Euler, $\partial T / \partial t \approx (T^{n+1} - T^n)/\Delta t$, the discretized form becomes:
\begin{equation}
\int_\Omega \rho_t c_p \frac{T^{n+1} - T^n}{\Delta t} w \, d\Omega
+ \int_\Omega k_t \nabla T^{n} \cdot \nabla w \, d\Omega
+ \int_\Omega \rho_b c_b \omega_b (T^{n} - T_b) w \, d\Omega
= \int_\Omega \sigma(T^{n}) |\nabla \phi|^2 w \, d\Omega.
\end{equation}





% \end{thebibliography}
\end{document}